\documentclass[reprint, superscriptaddress,amsmath,amssymb,aps]{revtex4-2}
\usepackage[utf8]{inputenc}

\usepackage{graphicx}
\usepackage{physics}
\usepackage[english]{babel}
\usepackage{dcolumn}
\usepackage{bm}
\usepackage[dvipsnames]{xcolor}
\usepackage{hyperref}
\usepackage[mathlines]{lineno}

\begin{document}

\preprint{APS/123-QED}

\title{Linear and nonlinear optical torque in multi-level atomic systems driven by counter-rotating orbital angular momentum fields}

\author{Dharma P. Permana}
\email{dharma.permana@ff.stud.vu.lt}
\email{dhp.permana@gmail.com}
\affiliation{%
 Institute of Theoretical Physics and Astronomy, Vilnius University, Saul\.{e}tekio 3, Vilnius LT-10257, Lithuania
}%
\affiliation{Laser Research Center, Vilnius University, Vilnius, LT-10223, Lithuania}
\affiliation{D\'{e}partement Physique, Facult\'{e} des Sciences, Aix-Marseille Universit\'{e}, Marseille, France}

\author{Mažena Mackoit Sinkevičienė}%
 \email{mazena.mackoit-sinkeviciene@ff.vu.lt}
\affiliation{%
 Institute of Theoretical Physics and Astronomy, Vilnius University, Saul\.{e}tekio 3, Vilnius LT-10257, Lithuania
}%

\author{Viačeslav Kudriašov}
\email{viaceslav.kudriasov@ff.vu.lt}
\affiliation{%
 Institute of Theoretical Physics and Astronomy, Vilnius University, Saul\.{e}tekio 3, Vilnius LT-10257, Lithuania
}

\author{Julius Ruseckas}
\email{julius.ruseckas@bpti.eu}
\affiliation{Baltic Institute of Advanced Technology, LT-01403 Vilnius, Lithuania}

\author{Gediminas Juzeli\"unas}%
 \email{gediminas.juzeliunas@tfai.vu.lt}
\affiliation{%
 Institute of Theoretical Physics and Astronomy, Vilnius University, Saul\.{e}tekio 3, Vilnius LT-10257, Lithuania
}%

\author{Hamid R. Hamedi}
 \email{hamid.hamedi@tfai.vu.lt}
\affiliation{%
 Institute of Theoretical Physics and Astronomy, Vilnius University, Saul\.{e}tekio 3, Vilnius LT-10257, Lithuania
}%

\date{\today}

\begin{abstract}
We investigate the generation of optical torque in coherently prepared multi-level atomic media driven by a vector vortex beam composed of two counter-rotating components carrying opposite orbital angular momenta, $+l\hbar$ and $-l\hbar$. We consider a three-level $\Lambda$ configuration and a four-level tripod configuration. Using a perturbative steady-state solution of the optical Bloch equations, we obtain analytical expressions for both linear and nonlinear contributions to the optical torque. The results show that the torque is strongly controlled by atomic coherence, including the initial population imbalance and the relative phase between the vortex components. Nonvanishing torque can arise even when the two components have equal amplitudes, due to coherence-induced asymmetry in the atomic response. In the tripod configuration, the presence of a strong control field leads to electromagnetically induced transparency, which suppresses the torque near resonance and shifts the dominant response to finite detunings. These results establish a route for controlling light-induced rotational dynamics in atomic media using vector vortex fields, with potential applications in coherent optical manipulation and angular-momentum-based control in quantum systems.
\end{abstract}

\maketitle


\section{\label{Introduction}Introduction}
Electromagnetically induced transparency (EIT) is one of the most important and intriguing effects arising from the interaction of light with an atomic medium, where quantum interference renders an initially opaque medium transparent under specific conditions \cite{boller1991,fleischhauer2000,Marangos.RMP2005}. Since the discovery and experimental demonstration of EIT, a wide range of novel phenomena have emerged. In particular, EIT enables slow-light propagation, in which the group velocity of a light pulse inside an atomic ensemble can be dramatically reduced to a few meters per second \cite{Hau1999,paspalakis2002,juzeliunas2004}, and paves the way for applications such as enhanced optical nonlinearities \cite{minxiao2001,gong2006,hamedi2015}, light storage \cite{fleischhauer2000,lukin2001,Hau2001}, and stationary light \cite{Otterbach2010,Peters2022,Kim2022}. 

The investigation of atom-light interaction in EIT-based configurations has been extended to structured light beams that carry orbital angular momentum (OAM), known as optical vortices \cite{coullet1989,allen1992}, which are characterized by a helical phase structure, $e^{il\phi}$, where the parameter $l$, known as the topological charge, is associated with the discrete OAM of the light beam, $l\hbar$ per photon. From this interaction, several interesting phenomena emerge, including the transfer of the OAM of light \cite{ruseckas2013,Hamid.PRA2019} and the entanglement of OAM states in photon pairs \cite{guo2008}. Going further, a full-vectorial treatment of optical vortices with spatially varying polarization states \cite{rosales2018} allows one to consider both the orbital angular momentum (OAM) and the spin angular momentum (SAM) of light, which are encoded in the spatial phase and polarization texture of the light beam, respectively \cite{zhang2010}. The interaction of vector vortices with EIT-based atomic media has led to the discovery of spatially dependent transparency \cite{radwell2017,tarak2017} and optical spin-orbit coupling, manifested as an exchange of polarization states controlled by the OAM of light \cite{Kudriasov.OE2025,Permana2025}. Furthermore, these effects, namely spatially dependent transparency and optical spin-orbit coupling, have also been theoretically demonstrated in the slow-light propagation regime \cite{Permana2026}, demonstrating the versatility of EIT-based atomic media for tailoring optical responses and enabling manipulation of the complex behavior of structured light beams.
\begin{figure*}[t!]
    \centering
    \includegraphics[width=1\textwidth]{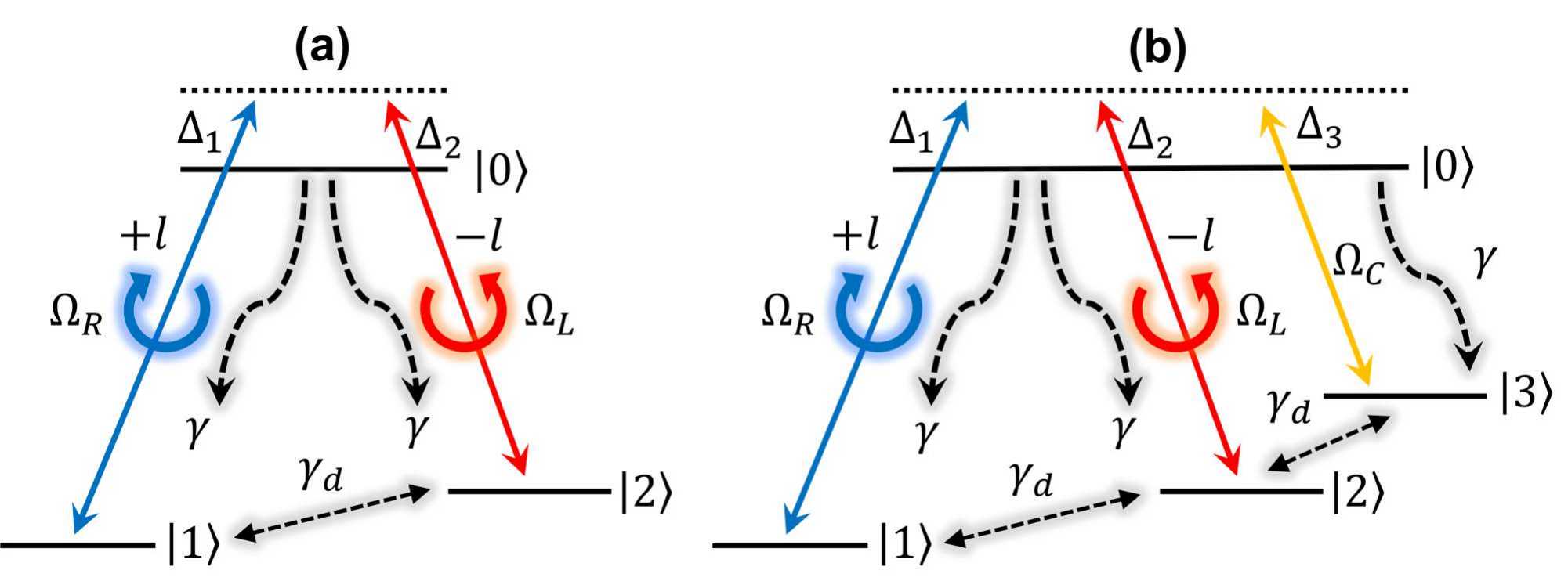}
    \caption{(a) Illustration of a three-level $\Lambda$-configuration atom–light coupling scheme. The right-handed circularly polarized vortex beam with Rabi amplitude $\Omega_{R}$ and OAM charge $+l$ couples the $\ket{1}\rightarrow\ket{0}$ transition, while the left-handed circularly polarized vortex beam with Rabi amplitude $\Omega_{L}$ and OAM charge $-l$ couples the $\ket{2}\rightarrow\ket{0}$ transition. (b) Illustration of a four-level tripod-configuration atom–light coupling scheme where an additional control field with Rabi amplitude $\Omega_C$ without OAM couples the $\ket{3}\rightarrow\ket{0}$ transition. The atomic systems in both configurations are assumed to undergo radiative and nonradiative relaxation processes with the decay rates $\gamma$ and $\gamma_d$, respectively.}
    \label{Lambda configuration}
\end{figure*}

Beyond optical responses, the interaction between a vortex beam and an atomic medium can also cause mechanical effects. In particular, the transfer of the OAM of light to the atoms induces a quantized torque on the medium \cite{Phillips2006}, which can be observed as rotational motion of atomic gases \cite{Babiker1994,Babiker2010}. The torque effect has been investigated in a simple three-level $\Lambda$ configuration interacting with two OAM beams, showing that the generated torque acting on the atomic medium can be controlled by the amplitudes and detunings of each beam \cite{Babiker2010}. Extending this investigation to a more complex five-level double-tripod (DT) configuration, where atoms interact with four strong control fields and two weak OAM probe beams, has revealed an additional degree of freedom for torque manipulation through coherent phase control \cite{Hamid2025}. However, previous studies have predominantly considered scalar vortex fields in which both optical components carry identical orbital angular momentum, resulting in effectively single-handed angular momentum transfer. In addition, although atomic coherence has been shown to play a crucial role in spin–orbit coupling and interference effects in coherently prepared atomic media \cite{Permana2025,Permana2026}, its impact on optical torque generation has not yet been fully explored.

In this work, we demonstrate that a vector vortex beam interacting with a coherently prepared multi-level atomic medium generates both linear and nonlinear optical torque contributions that are strongly sensitive to the initial atomic coherence. The driving field is constructed from two weak vortex components carrying opposite orbital angular momenta, $+l\hbar$ and $-l\hbar$, with orthogonal circular polarizations. These fields interact with a multi-level atom–light system prepared in a coherent superposition of two ground states (phaseonium) \cite{scully.book1997}. We first analyze a three-level $\Lambda$ configuration \cite{Permana2025}, driven solely by the weak vector vortex field, and then extend the model to a four-level tripod configuration \cite{Permana2026} incorporating an additional strong control field. Using a perturbative steady-state solution of the optical Bloch equations, we derive analytical expressions for both linear and nonlinear torque contributions. The results show that the torque is governed by the initial atomic coherence and can remain nonzero even when the counter-OAM components have equal amplitudes, due to coherence-induced asymmetry in the atomic response. This mechanism provides an additional degree of freedom for controlling angular momentum transfer in atomic media and may enable new schemes for coherent optical manipulation in quantum systems.

\section{\label{Theoretical Model}Theoretical Model of Light–Matter Interaction}

\subsection{\label{System}General description of the system}
We consider the interaction of a vector vortex beam with multi-level atomic systems. In particular, we investigate two configurations: a three-level $\Lambda$ system and a four-level tripod system, illustrated in Figs.~\ref{Lambda configuration}(a) and \ref{Lambda configuration}(b), respectively. The vector vortex beam consists of two weak optical components with Rabi frequencies $\Omega_R$ and $\Omega_L$. The field $\Omega_R$ is right-circularly polarized and carries an OAM charge of $+l$, while $\Omega_L$ is left-circularly polarized and carries an OAM charge of $-l$. 
In a three-level $\Lambda$ configuration, the fields $\Omega_R$ and $\Omega_L$ drive the transitions $\ket{1}\rightarrow\ket{0}$ and $\ket{2}\rightarrow\ket{0}$, respectively. The atom-light Hamiltonian under the dipole approximation and rotating-wave approximation is given by \cite{Permana2025}:
\begin{equation}
    H_\Lambda = -\hbar \sum_{j=1}^{2}\Delta_j\ket{j}\bra{j} + \hbar\Omega_{R} \ket{1}\bra{0} +\hbar\Omega_{L}\ket{2}\bra{0} + \mathrm{H.c.}, \label{eq.1}
\end{equation}
where the optical field detuning is defined as follows:
\begin{equation}
    \Delta_j=\omega_{0j}-\omega; \quad \omega_{0j}=\omega_0 - \omega_j, \label{eq.2}
\end{equation}
where $\omega$ denotes the optical frequency of both laser fields, while $\omega_j$ shows the eigen-frequency of the atomic bare states $\ket{j}$ with $j=1,2$.

In the four-level tripod configuration, a third strong laser field (control field) is introduced, with Rabi frequency $\Omega_C$. This field couples the additional ground state $\ket{3}$ with the excited state $\ket{0}$. The atom–light Hamiltonian under the dipole and rotating-wave approximations is given by \cite{Permana2026}:
\begin{equation}
\begin{aligned}
    H_{\mathrm{Tripod}} =& -\hbar \sum_{j=1}^{3}\Delta_j\ket{j}\bra{j} + \hbar\Omega_{R} \ket{1}\bra{0} \\
    &+\hbar\Omega_{L}\ket{2}\bra{0} +\hbar\Omega_C\ket{3}\bra{0} + \mathrm{H.c.}, \label{eq.13} 
\end{aligned}
\end{equation}
where the detunings of the weak probe fields are defined in Eq.~(\ref{eq.2}), while the detuning of the control field is given by
\begin{equation}
    \Delta_3 = \omega_{03} - \omega_C; \quad \omega_{03} = \omega_0 - \omega_3, \label{eq.14}
\end{equation}
with $\omega_C$ denotes the control field optical frequency, which may differ from the optical frequency of the two probes $\omega$, and $\omega_3$ denotes the eigen-frequency of the ground-state $\ket{3}$.

\subsection{\label{Torque}Optical Torque Induced by Vector Vortex Beams}
We consider that the right- and left-handed vortex fields counter-propagate along the $z$ axis, which is the optical axis of both fields. The fields $\Omega_{R}$ and $\Omega_{L}$ are written as
\begin{subequations}
\label{eq.21}
\begin{align}
\Omega_{R} &= \Omega_{R,0}e^{i\Phi_R(\textbf{R})}, \label{eq.21a} \\
\Omega_{L} &= \Omega_{L,0}e^{i\Phi_L(\textbf{R})}, \label{eq.21b}
\end{align}
\end{subequations}  
where $\Phi_{R(L)}(\textbf{R})$ denotes the spatial phase profile in terms of the center-of-mass coordinate $\mathbf{R}$. The phase structures are given by
\begin{subequations}
\label{eq.22}
\begin{align}
    \Phi_{R}(\textbf{R}) &= l\phi + kz, \label{eq.22a} \\
    \Phi_{L}(\textbf{R}) &= -l\phi -kz, \label{eq.22b}
\end{align}
\end{subequations}
where $k$ is the wave-number, $l$ is the OAM charge, and $\phi$ is the azimuthal angle of the transverse plane of the beam. The right- and left-handed amplitudes $\Omega_{R,0}$ and $\Omega_{L,0}$ are assumed to be spatially inhomogeneous and modeled using lowest-order Laguerre–Gaussian (LG) modes. Restricting the analysis to the beam waist at $z=0$, their transverse profiles are written as 
\begin{subequations}
\label{eq.23}
\begin{align}
    \Omega_{R,0} &= \varepsilon e^{i\varphi} \cos{(\alpha)} G(r), \label{eq.23a} \\
    \Omega_{L,0} &= \varepsilon \sin{(\alpha)} G(r), \label{eq.23b}
\end{align}
\end{subequations}
where $\varphi$ is the relative phase between the two polarization components, $\alpha$ controls their relative amplitude weights, and $\varepsilon$ is a real scaling constant. The radial envelope $G(r)$ is given by

\begin{equation}
    G(r) = \left(\frac{r}{w}\right)^{|l|}e^{-\frac{r^2}{w^2}}, \label{eq.24}
\end{equation}
where $r$ is the radial coordinate in the transverse plane and $w$ is the beam waist.

The expression for the optical force acting on the medium induced by the vector beam can be derived from the semiclassical Hamiltonians defined in Eqs.~(\ref{eq.1}) and (\ref{eq.13}) for the $\Lambda$ and tripod configurations, respectively. Since the optical force depends on the spatial gradient of the Hamiltonian, only terms containing spatial dependence contribute. In the present model, this dependence is fully contained in the spatial envelopes $\Omega_{R,0}$ and $\Omega_{L,0}$ for both configurations. Therefore, the relevant part of the Hamiltonian for the force calculation reduces to 
\begin{equation}
    H= \hbar\Omega_{R}\ket{1}\bra{0} + \hbar\Omega_{L}\ket{2}\bra{0} + \hbar\Omega_{R}^*\ket{0}\bra{1} + \hbar\Omega_{L}^*\ket{0}\bra{2}. \label{eq.25}
\end{equation}
The expectation value of this Hamiltonian is given by
\begin{equation}
    \left< H \right> = \hbar\Omega_{R}\sigma_{01} + \hbar\Omega_{R}^*\sigma_{10} + \hbar\Omega_{L}\sigma_{02} + \hbar \Omega_{L}^*\sigma_{20}, \label{eq.26}
\end{equation}
where $\sigma_{ij}=\langle i|\sigma|j\rangle$ denotes the elements of the density matrix in the laboratory frame. The coherences in the rotating-wave frame are related to $\sigma_{ij}$ as
\begin{subequations}
\label{eq.27}
\begin{align}
    \rho_{10} &= \sigma_{10}e^{-i\Phi_R(\textbf{R})}; \quad \rho_{01} = \sigma_{01}e^{i\Phi_R(\textbf{R})}, \label{eq.27a} \\
    \rho_{20} &= \sigma_{20}e^{-i\Phi_L(\textbf{R})}; \quad \rho_{02} = \sigma_{02}e^{i\Phi_L(\textbf{R})}. \label{eq.27b} 
\end{align}
\end{subequations}
The optical force can be derived from the Ehrenfest theorem as \cite{Cook1979}:
\begin{equation}
    \textbf{F} = \hbar\sigma_{01} \nabla\Omega_{R} + \hbar\sigma_{10} \nabla\Omega_{R}^* + \hbar\sigma_{02} \nabla\Omega_{L} + \hbar\sigma_{20} \nabla\Omega_{L}^*. \label{eq.28}
\end{equation}
Substituting Eq.~(\ref{eq.21}) into Eq.~(\ref{eq.28}), the optical force reduces to
\begin{equation}
    \textbf{F} = 2\hbar\mathrm{Im}\left[\Omega_{R,0}^*\rho_{10}\right]\nabla\Phi_R + 2\hbar\mathrm{Im}\left[\Omega_{L,0}^*\rho_{20}\right]\nabla\Phi_L, \label{eq.29}
\end{equation}
where the terms involving gradients of the field amplitudes (dipole force contributions) are neglected. This expression is obtained under the assumption that all atoms experience the same light-induced force, which corresponds to approximating the many-body wave function of the atomic ensemble as a product of identical single-atom wave functions \cite{Phillips2006}. Furthermore, the force acting on the center-of-mass coordinate $\mathbf{R}$ depends only on the elements of the  internal-state density matrix, assuming that the total wave function factorizes into the internal and center-of-mass components. 

The torque $\mathbf{T}$ exerted on the atomic center of mass around the beam axis can be obtained from the cross product between the radial component of $\mathbf{R}$ and the azimuthal component of the force $\mathbf{F}$, resulting in a torque vector directed along $\hat{z}$. Substituting the spatial phase profiles given in Eqs.~(\ref{eq.22a})–(\ref{eq.22b}) into Eq.~(\ref{eq.29}) and evaluating the gradient in cylindrical coordinates yields
\begin{subequations}
\label{eq.30}
\begin{align}
    \nabla\Phi_R &= \frac{l}{r}\hat{\phi} + k\hat{z}, \label{eq.30a} \\
    \nabla\Phi_L &= -\frac{l}{r}\hat{\phi} - k\hat{z}, \label{eq.30b}  
\end{align}
\end{subequations}
which leads to the force decomposition
\begin{equation}
\begin{aligned}
    \textbf{F} =& \frac{2\hbar l}{r}\left(\mathrm{Im}\left[\Omega_{R,0}^*\rho_{10}\right] - \mathrm{Im}\left[\Omega_{L,0}^*\rho_{20}\right] \right)\hat{\phi} \\
    &+2\hbar k\left(\mathrm{Im}\left[\Omega_{R,0}^*\rho_{10}\right] - \mathrm{Im}\left[\Omega_{L,0}^*\rho_{20}\right] \right)\hat{z}. \label{eq.31}
\end{aligned}
\end{equation}
Consequently, the vectorial light-induced torque along the propagation axis is obtained as
\begin{equation}
    \textbf{T} = \textbf{r}\times\textbf{F} = 2\hbar l\tau \hat{z}, \label{eq.32}
\end{equation}
which explicitly shows that the torque is quantized through the OAM charge $l$. Here $\tau$ is a scalar function determined by both the vector vortex field amplitudes $\Omega_{R,0}$, $\Omega_{L,0}$ and the atomic coherences $\rho_{10}$ and $\rho_{20}$, given by

\begin{equation}
    \tau = \mathrm{Im}\left[{\Omega}_{R,0}^*\rho_{10}\right] - \mathrm{Im}\left[{\Omega}_{L,0}^*\rho_{20}\right]. \label{eq.33}
\end{equation}

From the expression of the torque function in Eq.~(\ref{eq.33}), it follows that the torque exerted on the atomic system can remain non-vanishing even when the vortex pair carries opposite orbital angular momentum charges $+l$ and $-l$ in both the $\Lambda$ and tripod configurations. The resulting torque is both phase-dependent and coherence-dependent: it can be tuned through the relative-phase $\varphi$ between the right- and left-handed vortex amplitudes and through the atomic population parameter $\theta$. In Section~\ref{coherence torque spectra}, we analyze in detail the dependence of the torque on the relative phase of the vortex and the atomic populations in both the linear and nonlinear regimes. This is achieved by evaluating the coherences $\rho_{10}$ and $\rho_{20}$ for the $\Lambda$ and tripod configurations and separating their linear and nonlinear contributions.

To analyze the linear and nonlinear contributions to the torque function defined in Eq.~(\ref{eq.33}), we consider a perturbative expansion of the steady-state density matrix under the weak-field condition. The coherences can then be expressed as
\begin{equation}
    \rho_{ij} \approx \rho_{ij}^{(0)} + \rho_{ij}^{(1)} + \rho_{ij}^{(2)} + \rho_{ij}^{(3)} +\cdots, \label{eq.7}
\end{equation} 
where $\rho_{ij}^{(n)}$ denotes the $n$th-order correction in weak probe fields. The expansion preserves population normalization $\sum_j \rho_{jj}=1$. In this regime, the optical coherences relevant to torque, $\rho_{10}$ and $\rho_{20}$, acquire dominant contributions from the first- and third-order terms such that $\rho_{j0}\approx\rho_{j0}^{(1)}+\rho_{j0}^{(3)}$ $(j=1,2)$. Consequently, using the linearity of the imaginary part, the torque function in Eq.~(\ref{eq.33}) can be decomposed into linear and nonlinear contributions as
\begin{subequations}
\label{eq.34}
\begin{align}
\tau &= \tau^{(1)} + \tau^{(3)}, \label{eq.34a}\\
\tau^{(1)} &= \mathrm{Im}\left[{\Omega}_{R,0}^*\rho_{10}^{(1)}\right] - \mathrm{Im}\left[{\Omega}_{L,0}^*\rho_{20}^{(1)}\right], \label{eq.34b}\\
\tau^{(3)} &= \mathrm{Im}\left[{\Omega}_{R,0}^*\rho_{10}^{(3)}\right] - \mathrm{Im}\left[{\Omega}_{L,0}^*\rho_{20}^{(3)}\right]. \label{eq.34c}
\end{align}
\end{subequations}
From Eqs.~(\ref{eq.34b})–(\ref{eq.34c}), both the linear and nonlinear torque contributions, $\tau^{(1)}$ and $\tau^{(3)}$, depend on the radial coordinate in the transverse plane. This spatial dependence arises because the first- and third-order coherences are functions of the spatial envelopes $\Omega_{R,0}$ and $\Omega_{L,0}$ in both the $\Lambda$ and tripod configurations, as described by the solutions in Eqs.~(\ref{eq.9a})–(\ref{eq.9b}) and Eqs.~(\ref{eq.16a})–(\ref{eq.16b}) for the linear response, and in Eqs.~(\ref{eq.10a})–(\ref{eq.12b}) and Eqs.~(\ref{eq.18a})–(\ref{eq.20b}) for the nonlinear response.

Since both $\Omega_{R,0}$ and $\Omega_{L,0}$ share the radial Laguerre–Gaussian profile $G(r)$ defined in Eqs.~(\ref{eq.23a})–(\ref{eq.23b}), this spatial structure is directly inherited by the torque distribution across the transverse plane. As a result, the torque exhibits a ring-shaped spatial profile characteristic of the associated vector vortex beam. A detailed analysis of this radial dependence in both linear and nonlinear contributions is provided in Section~\ref{radially dependent torque}.

\subsection{\label{3-level system}Optical Bloch equations for the $\Lambda$ configuration}
The dynamics of the density matrix elements, including the optical coherences $\rho_{10}$ and $\rho_{20}$ associated with the probe-driven transitions, are governed by the Liouville–von Neumann equation (LVNE)
\begin{equation}
    \dot{\sigma} = -\frac{i}{\hbar}\left[H,\sigma\right] - \mathcal{L}_r \sigma - \mathcal{L}_c\sigma, \label{eq.3}
\end{equation}
where the atomic system is assumed to undergo both radiative and non-radiative relaxation processes, described by the superoperators \cite{tarak2017}:
\begin{subequations}
\label{eq.4}
\begin{align}
    \mathcal{L}_r\sigma&=\sum_{j=1}^{2}\frac{\gamma_{0j}}{2}\left(\ket{0}\bra{0}\sigma - \ket{j}\bra{j}\sigma_{00} + \sigma\ket{0}\bra{0} \right), \label{eq.4a}\\
    \mathcal{L}_c\sigma&=\sum_{j=1}^{2}\sum_{j\neq i =1}^{2}\frac{\gamma_{d}}{2}\left(\ket{j}\bra{j}\sigma - 2\ket{i}\bra{i}\sigma_{jj} + \sigma\ket{j}\bra{j} \right), \label{eq.4b}
\end{align}
\end{subequations}
where $\mathcal{L}_r$ and $\mathcal{L}_c$ describe the radiative and non-radiative relaxation processes, respectively. For cold atomic systems, the dephasing rate typically satisfies $\gamma_d \ll \gamma_{0j}$ \cite{tarak2017,Rebic2004}.

Assuming equal radiative decay rates, $\gamma_{0j}=\gamma$ ($j=1,2$), the optical Bloch equations take the form
\begin{subequations}
\label{eq.5}
\begin{equation}
    \dot{\rho}_{11} = i\left(\Omega_{R,0}^* \rho_{10} - \Omega_{R,0}\rho_{01} \right) + \gamma\rho_{00} +\gamma_d\left(\rho_{22} - \rho_{11} \right), \label{eq.5a}
\end{equation}

\begin{equation}
    \dot{\rho}_{22} = i\left(\Omega_{L,0}^* \rho_{20} - \Omega_{L,0}\rho_{02} \right) + \gamma\rho_{00} +\gamma_d\left(\rho_{11} - \rho_{22} \right), \label{eq.5b}
\end{equation}

\begin{equation}
\label{eq.5c}
    \begin{aligned}
        \dot{\rho}_{00} =& i\Omega_{R,0}\rho_{01}-i\Omega_{R,0}^* \rho_{10}  + i\Omega_{L,0}\rho_{02} - i\Omega_{L,0}^* \rho_{20} \\ 
        &+ 2\gamma\rho_{00}, 
    \end{aligned}
\end{equation}
\begin{equation}
\label{eq.5d} 
\begin{aligned}
    \dot{\rho}_{12} &= i(\Delta_1 - \Delta_2)\rho_{12} + i\Omega_{L,0}^*\rho_{10} - i\Omega_{R,0} \rho_{02} \\
    &- \gamma_d\rho_{12}, 
\end{aligned}
\end{equation}

\begin{equation}
\label{eq.5e} 
\begin{aligned}
    \dot{\rho}_{10} &= i\Delta_1\rho_{10} +i\Omega_{R,0}(\rho_{11} - \rho_{00}) + i\Omega_{L,0}\rho_{12}  \\ 
    &- \left(\gamma \rho_{10} + \frac{\gamma_d}{2}\right)\rho_{10}, 
\end{aligned}
\end{equation}

\begin{equation}
\label{eq.5f} 
\begin{aligned}
    \dot{\rho}_{20} &= i\Delta_2\rho_{20} +i\Omega_{L,0}(\rho_{22} - \rho_{00}) + i\Omega_{R,0}\rho_{21}  \\ 
    &- \left(\gamma \rho_{20} + \frac{\gamma_d}{2}\right)\rho_{20}. 
\end{aligned}
\end{equation}
\end{subequations}
The remaining equations follow from the Hermiticity condition $\rho_{ij}=\rho^*_{ji}$.

The atom is assumed to be initially prepared in a coherent superposition of the two ground states (phaseonium state), such that its wave function can be defined as   
\begin{equation}
    \ket{\psi(t=0)} = \cos{\theta} \ket{1} + \sin{\theta}\ket{2}, \label{eq.6}
\end{equation}
where $\theta$ denotes the coherence tuning angle. The coherence solutions are obtained using the perturbative expansion in Eq.~(\ref{eq.7}) together with the phaseonium condition in Eq.~(\ref{eq.6}). Under the weak-probe approximation, the excited-state population remains negligible in the steady-state regime. Therefore, the zeroth-order terms can be obtained as:
\begin{subequations}
\label{eq.8}
\begin{align}
    \rho_{00}^{(0)} &\approx 0, \label{eq.8a} \\
    \rho_{11}^{(0)} &\approx \cos^2{\theta}, \label{eq.8b} \\
    \rho_{22}^{(0)} &\approx \sin^{2}{\theta}, \label{eq.8c} \\
    \rho_{12}^{(0)} &= \rho_{21}^{(0)} \approx \sin{\theta}\cos{\theta}. \label{eq.8d} 
\end{align}
\end{subequations}
All other zero-order density-matrix elements vanish.

The steady-state solutions up to third order are obtained by substituting the expansion in Eq.~(\ref{eq.7}) into Eqs.~(\ref{eq.5a})–(\ref{eq.5f}) and imposing the steady-state condition $\dot{\rho}_{ij}=0$. The first-order solutions, which describe the linear optical response, are given by
\begin{subequations}
\label{eq.9}
\begin{align}
    \rho_{10}^{(1)} &\approx -\frac{\Omega_{R,0} \cos^2{\theta}+\Omega_{L,0}\sin{\theta}\cos{\theta}}{\Delta_1 + i\left(\gamma +\frac{\gamma_d}{2}\right)}, \label{eq.9a} \\
    \rho_{20}^{(1)} &\approx -\frac{\Omega_{R,0} \sin{\theta}\cos{\theta}+\Omega_{L,0}\sin^2{\theta}}{\Delta_2 + i\left(\gamma +\frac{\gamma_d}{2}\right)}, \label{eq.9b} \\
    \rho_{00}^{(1)} & = \rho_{11}^{(1)} = \rho_{22}^{(1)} \approx 0, \label{eq.9c} \\
    \rho_{12}^{(1)} & = \rho_{21}^{(1)} \approx 0. \label{eq.9d}
\end{align}        
\end{subequations}
The higher-order corrections responsible for the nonlinear response can be obtained recursively. For the second-order corrections, one obtains
\begin{subequations}
\label{eq.10}
\begin{align}
    \rho_{00}^{(2)} &\approx \frac{A+B}{2\gamma}, \label{eq.10a} \\
    \rho_{11}^{(2)} &\approx -\frac{\rho_{00}^{(2)}}{2} - \frac{A-B}{4\gamma_d}, \label{eq.10b} \\
    \rho_{22}^{(2)} &\approx -\frac{\rho_{00}^{(2)}}{2} + \frac{A-B}{4\gamma_d}, \label{eq.10c}  \\
    \rho_{12}^{(2)} &\approx \frac{\Omega_{R,0} \rho_{02}^{(1)}-\Omega_{L,0}^* \rho_{10}^{(1)}}{\Delta_1-\Delta_2+i\gamma_d}, \label{eq.10d} \\
    \rho_{10}^{(2)} &\approx \frac{\Omega_{R,0} \left( \rho_{00}^{(1)} - \rho_{11}^{(1)}\right)-\Omega_{L,0} \rho_{12}^{(1)}}{\Delta_1+i\left(\gamma + \frac{\gamma_d}{2}\right)}, \label{eq.10e} \\
    \rho_{20}^{(2)} &\approx \frac{-\Omega_{R,0} \rho_{21}^{(1)}+\Omega_{L,0} \left( \rho_{00}^{(1)} - \rho_{22}^{(1)}\right)}{\Delta_2+i\left(\gamma + \frac{\gamma_d}{2}\right)}, \label{eq.10f}
\end{align} 
\end{subequations}
where the auxiliary quantities $A$ and $B$ are defined by
\begin{subequations}
\label{eq.11}
\begin{align}
    A&= 2 \mathrm{Im}\left[\Omega_{R,0}^*\rho_{10}^{(1)}\right], \label{eq.11a} \\
    B&= 2 \mathrm{Im}\left[\Omega_{L,0}^*\rho_{20}^{(1)}\right]. \label{eq.11b}    
\end{align}  
\end{subequations}
At third order, we restrict our attention to $\rho_{10}^{(3)}$, $\rho_{20}^{(3)}$, and their complex conjugates, as these coherences directly determine both the optical susceptibility of the medium \cite{minxiao2001} and the light-induced torque generated by the weak probe transitions \cite{Hamid2025}. The corresponding third-order coherences satisfy the recursive relations
\begin{subequations}
\label{eq.12}
\begin{align}
    \rho_{10}^{(3)} &\approx \frac{\Omega_{R,0} \left( \rho_{00}^{(2)} - \rho_{11}^{(2)}\right)-\Omega_{L,0} \rho_{12}^{(2)}}{\Delta_1+i\left(\gamma + \frac{\gamma_d}{2}\right)}, \label{eq.12a} \\
    \rho_{20}^{(3)} &\approx \frac{-\Omega_{R,0} \rho_{21}^{(2)}+\Omega_{L,0} \left( \rho_{00}^{(2)} - \rho_{22}^{(2)}\right)}{\Delta_2+i\left(\gamma + \frac{\gamma_d}{2}\right)}. \label{eq.12b}
\end{align}
\end{subequations}

\subsection{\label{4-level system}Optical Bloch equations for the tripod configuration}
The tripod system is assumed to undergo the same radiative and non-radiative relaxation processes introduced for the $\Lambda$ configuration. The corresponding relaxation operators are obtained by extending the summations in Eqs.~(\ref{eq.4a})–(\ref{eq.4b}) to include the additional ground state $\ket{3}$ ($\sum_{j=1}^{2} \rightarrow\sum_{j=1}^{3}$). Using the LVNE in Eq.~(\ref{eq.3}), together with the tripod Hamiltonian in Eq.~(\ref{eq.13}), and assuming equal radiative decay rates $\gamma_{0j}=\gamma$ ($j=1,2,3$), the optical Bloch equations take the form
\begin{subequations}
\label{eq.14}
\begin{equation}
\begin{aligned}
    \dot{\rho}_{00} =& i(\Omega_{R,0}\rho_{01} - \Omega_{R,0}^*\rho_{10}+\Omega_{L,0}\rho_{02}-\Omega_{L,0}^*\rho_{20})\\ &+ i(\Omega_C\rho_{03} - \Omega_C^*\rho_{30}) - 3\gamma\rho_{00}, \label{eq.15a}
\end{aligned}
\end{equation}

\begin{equation}
\begin{aligned}
\dot{\rho}_{11} = &i(\Omega_{R,0}^*\rho_{10}-\Omega_{R,0}\rho_{01}) + \frac{\gamma}{2}\rho_{00} \\ 
&+ \gamma_d(\rho_{22}+\rho_{33}-2\rho_{11}), \label{eq.15b} 
\end{aligned}
\end{equation}

\begin{equation}
\begin{aligned}
\dot{\rho}_{22} = &i(\Omega_{L,0}^*\rho_{20}-\Omega_{L,0}\rho_{02}) + \frac{\gamma}{2}\rho_{00} \\ &+ \gamma_d(\rho_{11}+\rho_{33}-2\rho_{22}), \label{eq.15c}
\end{aligned}
\end{equation}

\begin{equation}
\begin{aligned}
    \dot{\rho}_{33} = &i(\Omega_C^*\rho_{30}-\Omega_C\rho_{03}) + \frac{\gamma}{2}\rho_{00} \\ &+ \gamma_d(\rho_{11}+\rho_{22}-2\rho_{33}), \label{eq.15d}
\end{aligned}
\end{equation}

\begin{equation}
\begin{aligned}
    \dot{\rho}_{12} = &i(\rho_{12}(\Delta_1 - \Delta_2)+\Omega_{L,0}^*\rho_{10}-\Omega_{R,0}\rho_{02}) \\ &- 2 \gamma_d \rho_{12}, \label{eq.15e}
\end{aligned}
\end{equation}

\begin{equation}
\begin{aligned}
    \dot{\rho}_{13} = &i(\rho_{13}(\Delta_1 - \Delta_C) + \Omega_C^*\rho_{10}-\Omega_{R,0}\rho_{03}) \\ &- 2\gamma_d\rho_{13}, \label{eq.15f}
\end{aligned}
\end{equation}

\begin{equation}
\begin{aligned}
    \dot{\rho}_{10} = &i(\Omega_{R,0} \rho_{11} + \Omega_{L,0} \rho_{12} + \Omega_C \rho_{13} + \Delta_1\rho_{10})\\ &- i\Omega_{R,0} \rho_{00} - \frac{3}{2}\gamma\rho_{10} - \gamma_d\rho_{10}, \label{eq.15g}
\end{aligned}
\end{equation}

\begin{equation}
\begin{aligned}
    \dot{\rho}_{23} = &i(\rho_{23}(\Delta_2 - \Delta_C) + \Omega_C^*\rho_{20}-\Omega_{L,0}\rho_{03}) \\ &- 2\gamma_d\rho_{23}, \label{eq.15h}
\end{aligned}
\end{equation}

\begin{equation}
\begin{aligned}
    \dot{\rho}_{20} = &i(\Omega_{R,0} \rho_{21} + \Omega_{L,0} \rho_{22} + \Omega_C \rho_{23} + \Delta_2\rho_{20})\\ &- i\Omega_{L,0} \rho_{00} - \frac{3}{2}\gamma\rho_{20} - \gamma_d\rho_{20}, \label{eq.15i}
\end{aligned}
\end{equation}

\begin{equation}
\begin{aligned}
    \dot{\rho}_{30} = &i(\Omega_{R,0} \rho_{31} + \Omega_{L,0} \rho_{32} + \Omega_C \rho_{33} + \Delta_C\rho_{30})\\ &- i\Omega_C \rho_{00} - \frac{3}{2}\gamma\rho_{30} - \gamma_d\rho_{30}. \label{eq.15j}
\end{aligned}
\end{equation}
\end{subequations}
The remaining equations follow from the Hermiticity condition $\rho_{ij}=\rho_{ji}^{*}$ together with the normalization condition $\sum_j\rho_{jj}=1$.

The steady-state solutions are obtained in the weak-probe regime $|\Omega_{R,0}|,|\Omega_{L,0}|\ll|\Omega_C|$, by performing the perturbative expansion defined in Eq.~(\ref{eq.7}) with respect to the probe fields. We further assume the same phaseonium preparation as in Eq.~(\ref{eq.6}), with the additional ground state $\ket{3}$ initially unoccupied. Under these conditions, the zeroth-order density-matrix elements are identical to those of the $\Lambda$ configuration and are given by Eqs.~(\ref{eq.8a})–(\ref{eq.8d}), while all remaining zeroth-order terms vanish. 

Proceeding as in the $\Lambda$ configuration, the coherence solutions can be obtained recursively up to the third order. The first-order coherences are given by 
\begin{subequations}
\label{eq.15}
\begin{align}
    \rho_{10}^{(1)} &\approx -\frac{\Omega_{R,0} \cos^2{\theta}+\Omega_{L,0}\sin{\theta}\cos{\theta}}{\delta_1-\frac{|\Omega_C|^2}{\delta_{13}}}, \label{eq.16a} \\
    \rho_{20}^{(1)} &\approx -\frac{\Omega_{R,0} \sin{\theta}\cos{\theta}+\Omega_{L,0}\sin^2{\theta}}{\delta_2-\frac{|\Omega_C|^2}{\delta_{23}}}, \label{eq.16b} 
\end{align}        
\end{subequations}
where $\delta_j$ and $\delta_{ij}$ for $i\neq j=1,2,3$, are defined as:
\begin{subequations}
\label{eq.17}
\begin{align}
    \delta_j &= \Delta_j + i\Gamma, \label{eq.17a} \\
    \delta_{ij} &= \Delta_i-\Delta_j+2i\gamma_d, \label{eq.17b} 
\end{align}
\end{subequations}
with $\Gamma=\frac{3}{2}\gamma+\gamma_d$. All other first-order coherences vanish under the phaseonium preparation defined in Eq.~(\ref{eq.6}). The non-vanishing second-order corrections are
\begin{subequations}
\label{eq.18}
\begin{align}
\rho_{00}^{(2)}&=\frac{\left(1+\frac{2\Gamma|\Omega_C|^2}{3\gamma_d|\delta_3|^2}\right)C-2i \mathrm{Im}\left[iD^*\Omega_C^*\right]}{3\gamma+\frac{2\Gamma|\Omega_C|^2}{3\gamma_d|\delta_3|^2}\left(\gamma+4\gamma_d\right)}, \label{eq.18a} \\
    \rho_{11}^{(2)}&= \frac{\rho_{00}^{(2)}(\frac{1}{2}\gamma-\gamma_d)-A}{3\gamma_d}, \label{eq.18b}\\
    \rho_{22}^{(2)}&= \frac{\rho_{00}^{(2)}(\frac{1}{2}\gamma-\gamma_d)-B}{3\gamma_d}, \label{eq.18c} \\
    \rho_{33}^{(2)}&= -\rho_{00}^{(2)}-\rho_{11}^{(2)}-\rho_{22}^{(2)}, \label{eq.18d} \\
    \rho_{30}^{(2)} &= \frac{\Omega_c}{\delta_3}\left(\frac{\gamma+4\gamma_d}{3\gamma_d}\rho_{00}^{(2)}-\frac{C}{3\gamma_d}\right) + D^*, \label{eq.18e} \\
    \rho_{12}^{(2)} &= \frac{\Omega_{R,0}\rho_{02}^{(1)}-\Omega_{L,0}^*\rho_{10}^{(1)}}{\delta_{12}}, \label{eq.18f} 
\end{align}
\end{subequations}
where $A$ and $B$ are defined in Eqs.~(\ref{eq.11a})-(\ref{eq.11b}), while the auxiliary quantities $C$ and $D$ are defined by
\begin{subequations}
\label{eq.19}
\begin{align}
    C& = A+B, \label{eq.19a} \\
    D& = i\Omega_{R,0}^*\rho_{31}^{(1)} + i\Omega_{L,0}^* \rho_{32}^{(1)}. \label{eq.19b}
\end{align}
\end{subequations}
Finally, retaining only the third-order coherences associated with the vortex-driven probe transitions, one obtains
\begin{subequations}
\label{eq.20}
\begin{align}
    \rho_{10}^{(3)}&= \frac{\Omega_{R,0}(\rho_{00}^{(2)}-\rho_{11}^{(2)})-\Omega_{L,0}\rho_{12}^{(2)}}{\delta_1-\frac{|\Omega_C|^2}{\delta_{13}}} - \frac{\Omega_C\Omega_{R,0}\rho_{03}^{(2)}}{\delta_1\delta_{13}-|\Omega_C|^2}, \label{eq.20a} \\
    \rho_{20}^{(3)}&= \frac{\Omega_{L,0}(\rho_{00}^{(2)}-\rho_{22}^{(2)})-\Omega_{R,0}\rho_{21}^{(2)}}{\delta_2-\frac{|\Omega_C|^2}{\delta_{23}}} - \frac{\Omega_C\Omega_{L,0}\rho_{03}^{(2)}}{\delta_2\delta_{23}-|\Omega_C|^2} \label{eq.20b}.    
\end{align}
\end{subequations}

\section{\label{coherence sensitive torque}Linear and Nonlinear Contributions to the Optical Torque}
\subsection{\label{radially dependent torque}Radial Structure of the Optical Torque}
Equation~(\ref{eq.32}) shows that the spatial distribution of the optical torque is determined by the torque function $\tau$, which depends on the probe vortex amplitudes $\Omega_{R,0}$ and $\Omega_{L,0}$ as well as on the atomic coherences $\rho_{10}$ and $\rho_{20}$. Consequently, the torque distribution is governed jointly by the spatial structure of the vector vortex field and the coherence properties of the atomic medium. From the analytical solutions obtained for the $\Lambda$ and tripod configurations, it follows that the coherences inherit the radial dependence of the probe vortex amplitudes through the Laguerre--Gaussian profile defined in Eqs.~(\ref{eq.23a})--(\ref{eq.24}). As a result, the optical torque exhibits a nonuniform radial distribution across the transverse plane. In what follows, we investigate the radial structure of the optical torque in both configurations. For clarity, we restrict our analysis to the resonant condition $\Delta_1=\Delta_2=\Delta_3=0$ and consider vector vortex beams carrying an OAM charge of $|l|=1$. 
\begin{figure}[!h]
    \centering
    \includegraphics[width=1\linewidth]{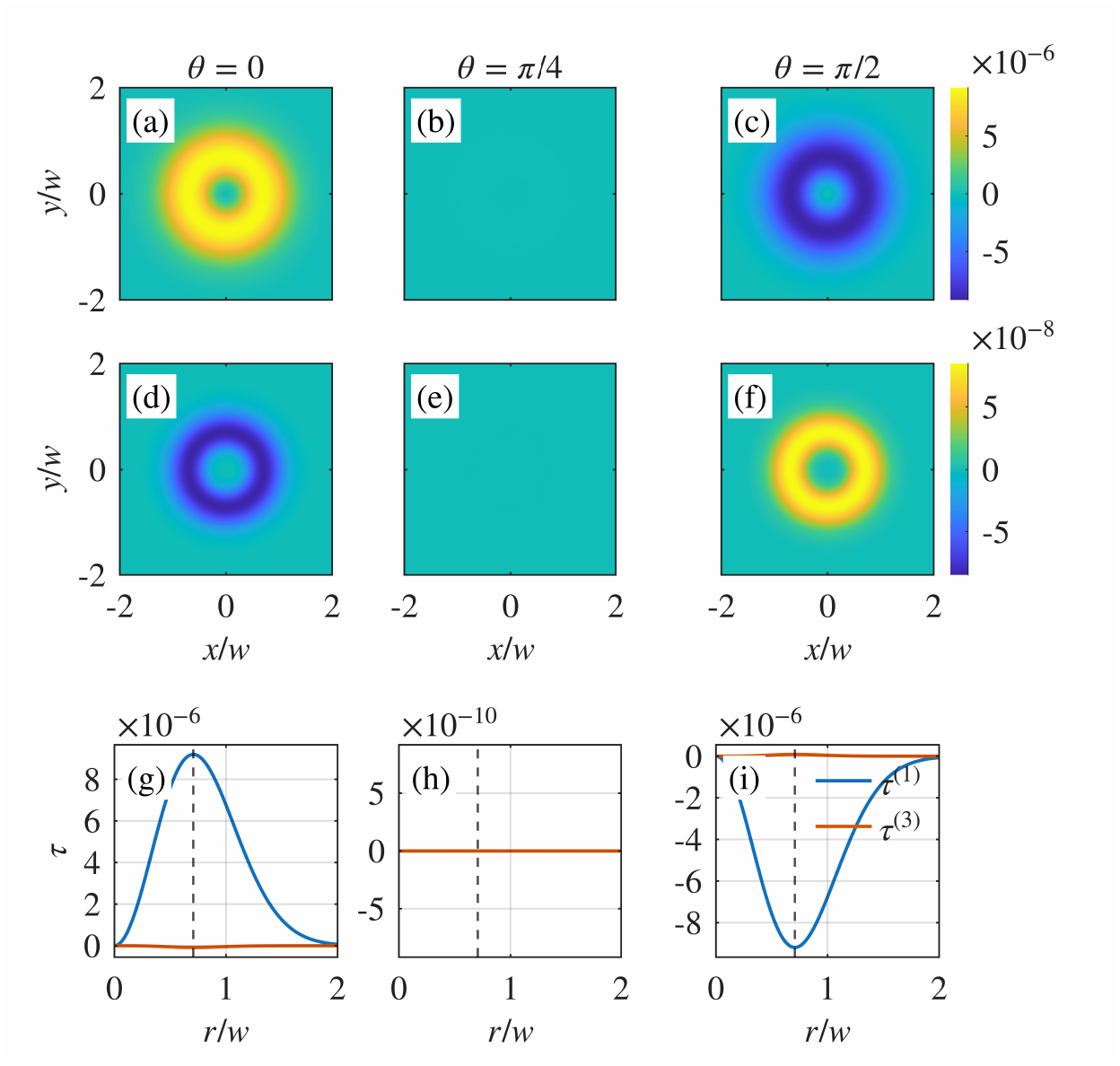}
    \caption{Spatial distribution of the optical torque in the $\Lambda$ system for $|l|=1$ under resonant conditions ($\Delta_1=\Delta_2=0$). (a)–(c) Linear contribution $\tau^{(1)}$ to the torque. (d)–(f) Third-order nonlinear contribution $\tau^{(3)}$ to the torque. (g)–(i) Radial dependence of both linear and nonlinear torque contributions at a fixed azimuthal angle $\phi=0$. The remaining parameters are $\varepsilon=10^{-2}\gamma$ and $\gamma_d=10^{-3}\gamma$. The transverse coordinates $x$, $y$, and $r$ are normalized to the Gaussian beam waist $w$. The radial position of maximum torque intensity is $r_0 = w/\sqrt{2}$.}
    \label{Torque radial pattern in Lambda}
\end{figure}

\begin{figure}[!h]
    \centering
    \includegraphics[width=1\linewidth]{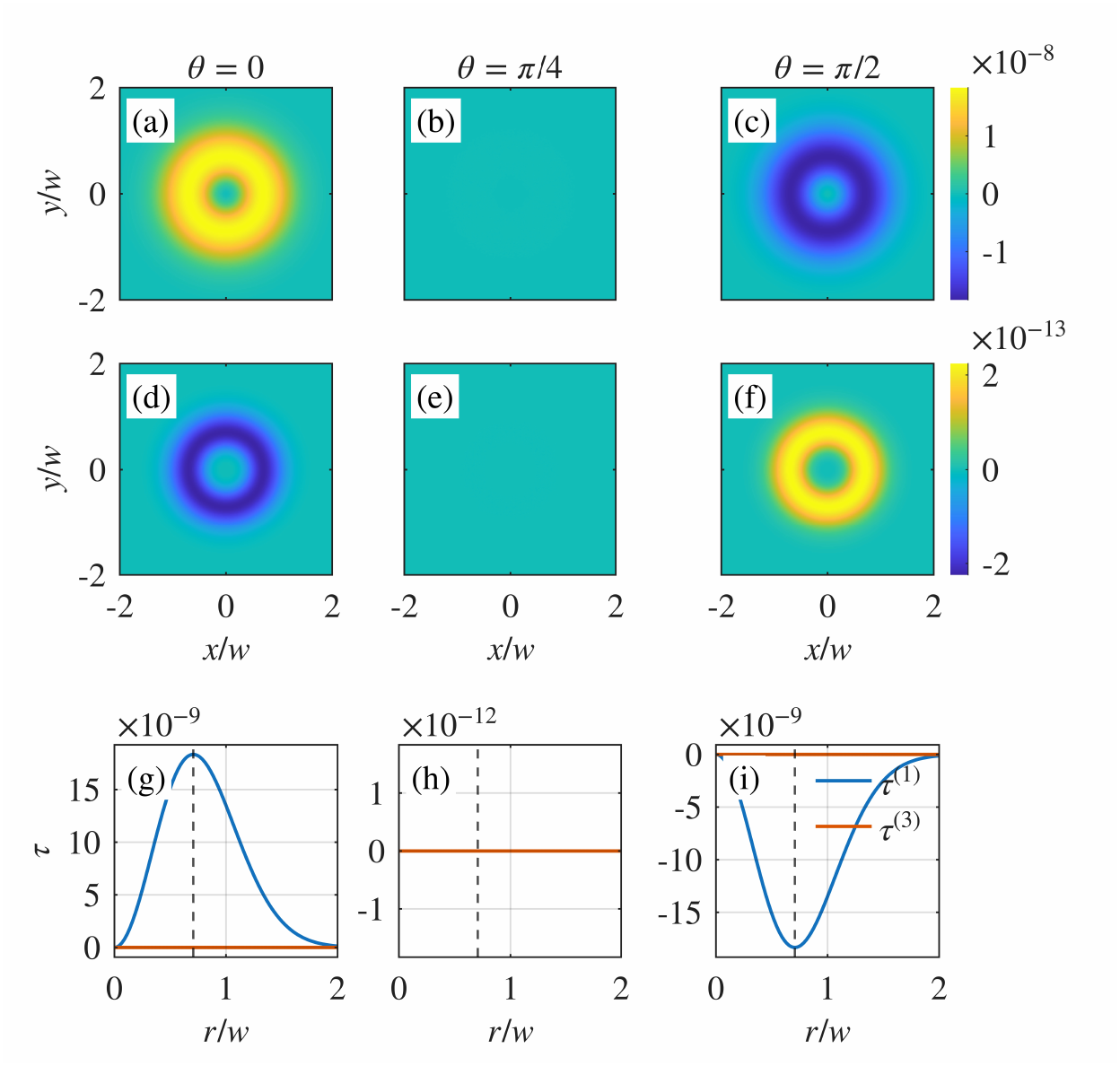}
    \caption{Spatial distribution of the optical torque in the tripod system for $|l|=1$ under resonant conditions ($\Delta_1=\Delta_2=\Delta_3=0$). (a)–(c) Linear contribution $\tau^{(1)}$ to the torque. (d)–(f) Third-order nonlinear contribution $\tau^{(3)}$ to the torque. (g)–(i) Radial dependence of both linear and nonlinear torque contributions at a fixed azimuthal angle $\phi=0$. The remaining parameters are $\varepsilon=10^{-2}\gamma$, $\gamma_d=10^{-3}\gamma$, and $|\Omega_C|=\gamma$. The transverse coordinates $x$, $y$, and $r$ are normalized to the Gaussian beam waist $w$. The radial position of maximum torque intensity is $r_0 = w/\sqrt{2}$.}
    \label{Torque radial pattern in Tripod}
\end{figure}
First, we consider the case where the control field $\Omega_C$ is absent, corresponding to the $\Lambda$ configuration. For clarity, we set the amplitude-tuning parameter $\alpha=\pi/4$ and the relative phase $\varphi=0$, so that the amplitudes of the right- and left-handed beam  are equal, $\Omega_{R,0}=\Omega_{L,0}$. In this regime, the optical torque is fully determined by the atomic coherences. Figure~\ref{Torque radial pattern in Lambda} shows the spatial distribution of the torque in the $\Lambda$ system for different initial atomic population configurations, $\theta=0$, $\pi/4$, and $\pi/2$.
It is observed that the torque forms a ring-shaped distribution across the transverse plane in both the linear regime [Figs.~\ref{Torque radial pattern in Lambda}(a)–\ref{Torque radial pattern in Lambda}(c)] and the nonlinear regime [Figs.~\ref{Torque radial pattern in Lambda}(d)–\ref{Torque radial pattern in Lambda}(f)]. The torque profiles are invariant with respect to the azimuthal angle $\phi$, while the radial dependence is inherited from the LG envelope $G(r)$ defined in Eq.~(\ref{eq.23}). As a result, the spatial structure of the optical torque is governed by the radial mode profile of the beam. The radial position at which the torque reaches its maximum depends on the OAM charge $|l|$ and can be obtained from Eq.~(\ref{eq.23}) as $r_0 = w\sqrt{\frac{|l|}{2}}$. Figures~\ref{Torque radial pattern in Lambda}(g)–\ref{Torque radial pattern in Lambda}(i) show the one-dimensional radial dependence of both linear and nonlinear contributions at $\phi=0$, where the maximum occurs at $r_0 = \frac{w}{\sqrt{2}}$ for $|l|=1$.

\begin{figure*}[t!]
    \centering
    \includegraphics[width=1\textwidth]{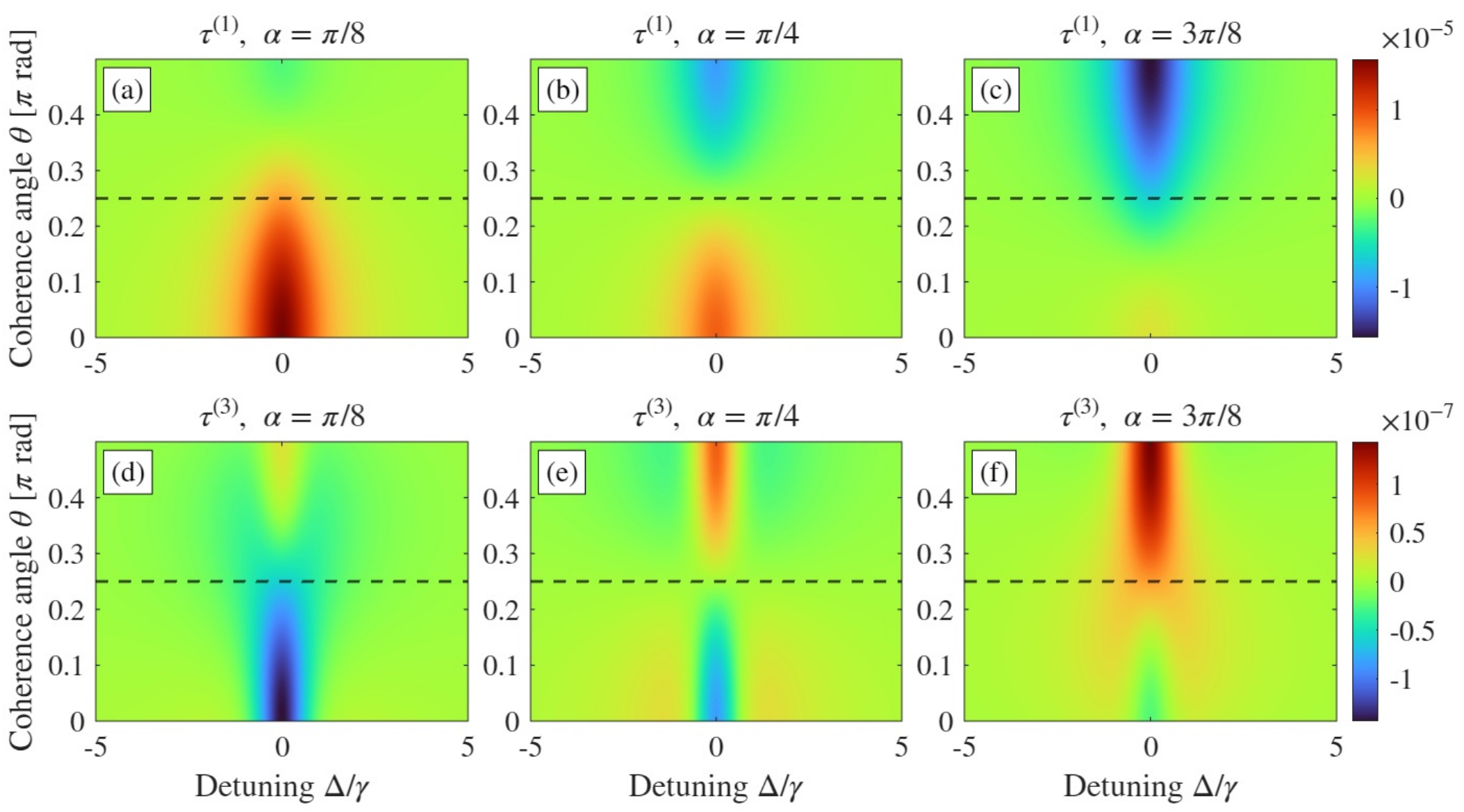}
    \caption{Parameter maps of the optical torque in the $\Lambda$ configuration for $|l|=1$ at fixed relative phase $\varphi=0$. (a)–(c) Linear contribution to the torque. (d)–(f) Nonlinear contribution to the torque. Three values of the relative amplitude parameter $\alpha$ are considered: (a),(d) $\alpha=\pi/8$; (b),(e) $\alpha=\pi/4$; (c),(f) $\alpha=3\pi/8$. The vertical axis represents the coherence tuning angle $\theta$, where $\theta=\pi/4$ is indicated by a horizontal dashed line. The horizontal axis corresponds to the normalized two-photon detuning. The remaining parameters are $\varepsilon=10^{-2}\gamma$ and $\gamma_d=10^{-3}\gamma$.}
    \label{Torque spectra lambda alpha variation}
\end{figure*}

\begin{figure*}[!t]
    \centering
    \includegraphics[width=1\textwidth]{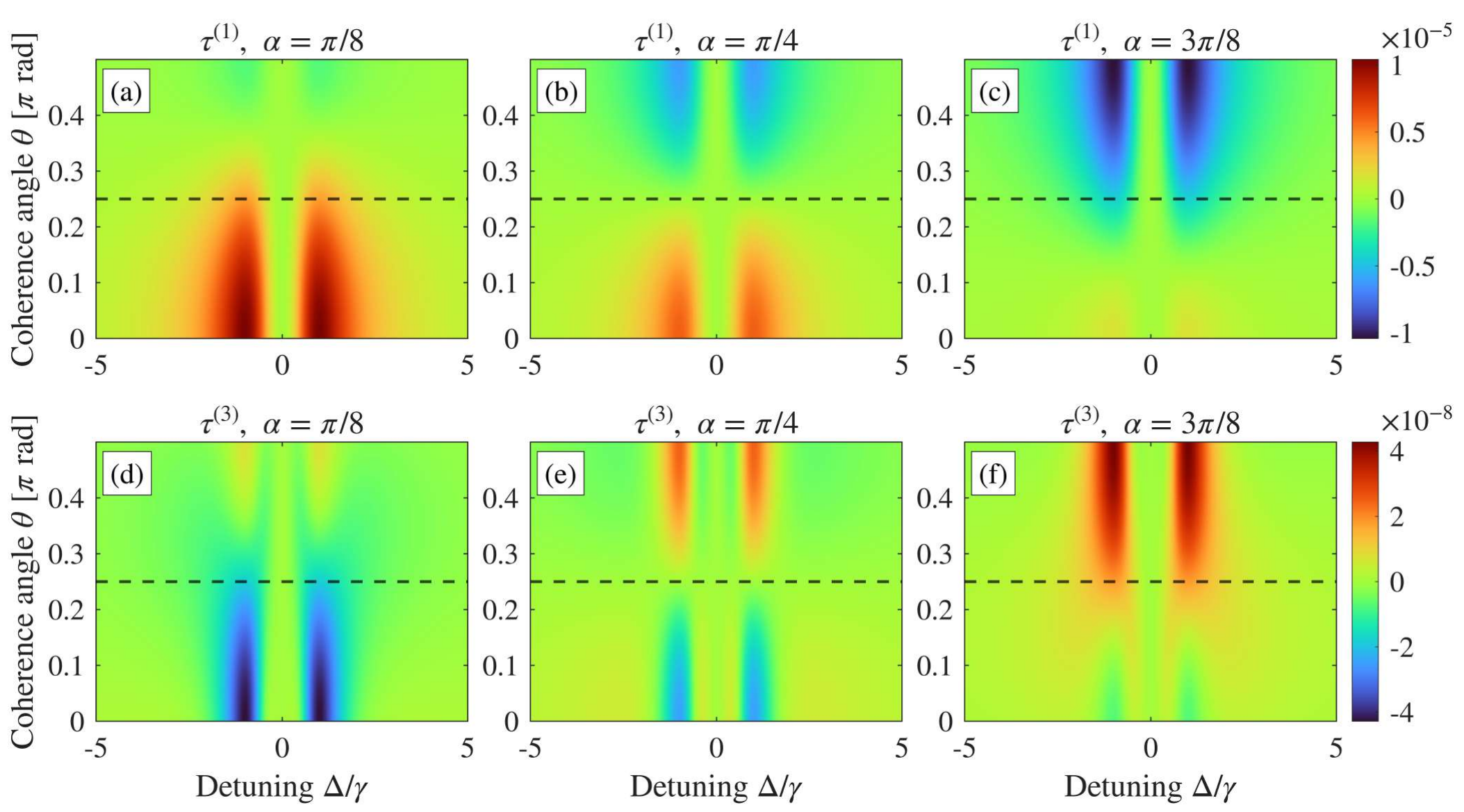}
    \caption{Parameter maps of the optical torque in the tripod configuration for $|l|=1$ at fixed relative phase $\varphi=0$. (a)–(c) Linear contribution to the torque. (d)–(f) Nonlinear contribution to the torque. Three values of the relative amplitude parameter $\alpha$ are considered: (a),(d) $\alpha=\pi/8$; (b),(e) $\alpha=\pi/4$; (c),(f) $\alpha=3\pi/8$. The vertical axis corresponds to the coherence tuning angle $\theta$, where $\theta=\pi/4$ is indicated by a horizontal dashed line. The horizontal axis corresponds to the normalized two-photon detuning. The remaining parameters are $\varepsilon=10^{-2}\gamma$, $\gamma_d=10^{-3}\gamma$, and $|\Omega_C|=\gamma$.}
    \label{Torque spectra tripod alpha variation}
\end{figure*}

The torque profiles shown in Figs.~\ref{Torque radial pattern in Lambda}(a)–(i) highlight the role of atomic coherence in controlling the vector-vortex-induced optical torque. Although the applied probe field is a vector vortex beam composed of two counter-rotating OAM components carrying $+\hbar l$ and $-\hbar l$, the resulting optical torque generally does not cancel and is strongly dependent on the initial atomic populations. In both linear and nonlinear regimes, the torque vanishes only for $\theta=\pi/4$, where the two ground states $\ket{1}$ and $\ket{2}$ are equally populated, as shown in Figs.~\ref{Torque radial pattern in Lambda}(b), (e), and (h). In this case, the system forms a coherent population trapping (CPT) state, in which the atoms are driven into a dark superposition and the medium becomes transparent to the vector vortex probe field \cite{Permana2025}. For $\theta=0$, where the population is entirely in $\ket{1}$, both linear and nonlinear contributions to the optical torque are nonzero and reach their maximum values [Figs.~\ref{Torque radial pattern in Lambda}(a), (d), and (g)]. However, the nonlinear contribution exhibits opposite sign and is suppressed by approximately two orders of magnitude compared to the linear response for $\gamma_d=10^{-3}\gamma$. For $\theta=\pi/2$, where the whole population is initially in $\ket{2}$, the torque exhibits a similar behavior but with a reversed sign, as shown in Figs.~\ref{Torque radial pattern in Lambda}(c), (f), and (i). This symmetry reflects the role of ground-state population imbalance in determining the direction of the vector-vortex-induced optical torque. 

Next, we consider the case where the control field is present, which corresponds to the tripod configuration. In this system, the control field strength is fixed at $|\Omega_C|=\gamma$, while all other parameters are kept identical to those used in the $\Lambda$ configuration (see Fig.~\ref{Torque radial pattern in Lambda}). The resulting torque profiles in both the linear and nonlinear regimes retain the ring-shaped spatial structure governed by the radial function $G(r)$, with the same radial position of maximum intensity as in the $\Lambda$ system, as shown in Fig.~\ref{Torque radial pattern in Tripod}. The optical torque in the tripod configuration also exhibits a pronounced dependence on the coherence parameter $\theta$, for both linear and nonlinear contributions, similar to the $\Lambda$ case. However, a key distinction arises in the presence of the strong control field. Although the qualitative structure of the torque distribution remains unchanged, the magnitude of the linear contribution is reduced by approximately two orders of magnitude, and the nonlinear contribution is suppressed by about five orders of magnitude compared with the $\Lambda$ configuration under resonant conditions ($\Delta_1=\Delta_2=\Delta_3=0$). This strong suppression is a manifestation of EIT, which emerges due to the control field $\Omega_C$ and renders the medium nearly transparent to the weak vector vortex probe fields $\Omega_{R,0}$ and $\Omega_{L,0}$ \cite{Permana2026}. Consequently, the optical torque generated by the structured light field is significantly reduced in the tripod configuration despite the preserved spatial structure of the vector vortex beam. 

\subsection{\label{coherence torque spectra}Coherence dependent torque}
In this section, we analyze the dependence of the optical torque on the normalized detuning $\Delta/\gamma$ for different atomic ground-state preparations, characterized by the coherence tuning angle $\theta$. We show that the torque response in both the linear and nonlinear regimes, for both configurations, is strongly influenced by the properties of the vector vortex probe field, namely the relative amplitude parameter $\alpha$ and the relative phase $\varphi$. For clarity, we restrict the analysis to the case $|l|=1$ and evaluate the torque at the radial position of maximum intensity, $r_0 = w/\sqrt{2}$. Due to the azimuthal symmetry of the torque distribution, we further fix the azimuthal angle at $\phi=0$ without loss of generality. The remaining parameters, including the amplitude scaling factor $\varepsilon$ and the dephasing rate $\gamma_d$, are the same as those used in Section~\ref{radially dependent torque}.

First, we consider the $\Lambda$ configuration in the absence of the control field. Figure~\ref{Torque spectra lambda alpha variation} shows the dependence of the optical torque on the normalized detuning $\Delta/\gamma$ for different values of the coherence tuning angle $\theta$, with the relative phase between the right- and left-handed vortex components fixed at $\varphi=0$. It is observed that both the linear and nonlinear contributions to the torque, and their dependence on populations of the ground-state through $\theta$, are modified when the relative amplitude parameter $\alpha$ is varied. However, for all values of $\alpha$, a nonzero torque is maintained in the vicinity of resonance $\Delta/\gamma \approx 0$.

Figures~\ref{Torque spectra lambda alpha variation}(b) and \ref{Torque spectra lambda alpha variation}(e) correspond to the balanced case $\alpha=\pi/4$, for which the right- and left-handed components of the incident vector vortex field have equal amplitudes, resulting in an effectively linearly polarized input field. In this case, the torque vanishes at $\theta=\pi/4$, where the two ground states are equally populated. Away from this condition, both the linear and nonlinear contributions become nonzero. For $0 \leq \theta < \pi/4$, where the population of $\ket{1}$ exceeds that of $\ket{2}$, the linear torque is positive, corresponding to a rotation along $+\hat{z}$, whereas the nonlinear contribution has the opposite sign. For $\pi/4 < \theta \leq \pi/2$, where $\ket{2}$ becomes more populated than $\ket{1}$, both contributions reverse sign, producing a torque of the same magnitude but opposite direction compared to the case $0 \leq \theta < \pi/4$. This behavior reflects the coherence-sensitive nature of the torque and demonstrates that its sign can be controlled by the relative population of the two ground states.

\begin{figure*}[!t]
    \centering
    \includegraphics[width=1\textwidth]{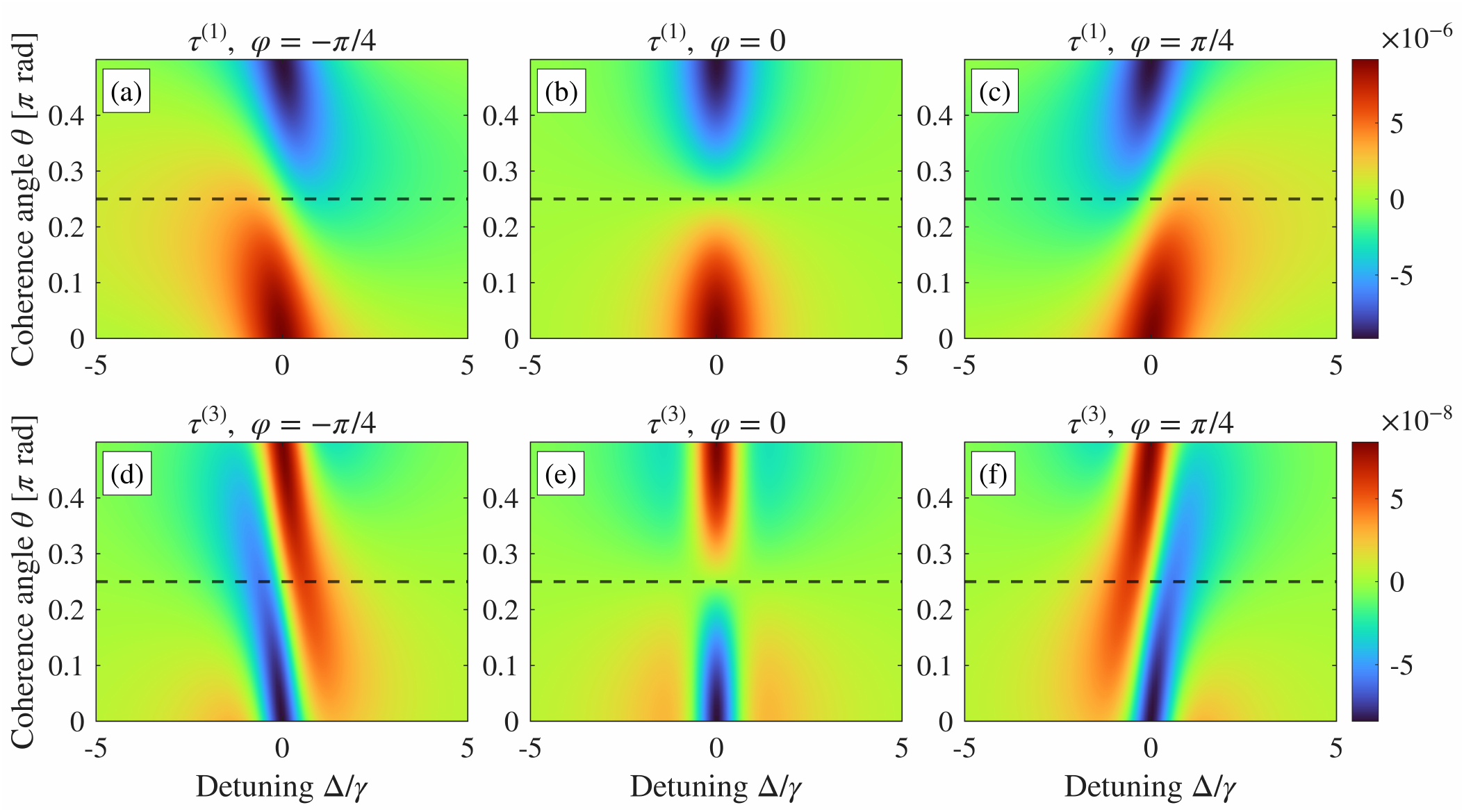}
    \caption{Parameter maps of the optical torque in the $\Lambda$ configuration for $|l|=1$ at fixed relative amplitude strength $\alpha=\pi/4$. (a)–(c) Linear contribution to the torque. (d)–(f) Nonlinear contribution to the torque. Three cases of the relative phase $\varphi$ are considered: (a),(d) $\varphi=-\pi/4$; (b),(e) $\varphi=0$; (c),(f) $\varphi=\pi/4$. The vertical axis corresponds to the coherence tuning angle $\theta$, where $\theta=\pi/4$ is indicated by a horizontal dashed line. The horizontal axis corresponds to the normalized two-photon detuning. The remaining parameters are $\varepsilon=10^{-2}\gamma$ and $\gamma_d=10^{-3}\gamma$.}
    \label{Torque spectra lambda varphi}
\end{figure*}

 In the regime $0\leq\alpha<\pi/4$, the incident vector vortex beam is dominated by its right-handed component polarized circularly, since $|\Omega_{R,0}|>|\Omega_{L,0}|$. The corresponding linear and nonlinear torque maps for $\alpha=\pi/8$ are shown in Figs.~\ref{Torque spectra lambda alpha variation}(a) and \ref{Torque spectra lambda alpha variation}(d), respectively. For the linear contribution shown in Fig.~\ref{Torque spectra lambda alpha variation}(a), the torque is predominantly positive, corresponding to a rotation directed along $+\hat{z}$. In contrast, the nonlinear contribution in Fig.~\ref{Torque spectra lambda alpha variation}(d) is predominantly negative, corresponding to a torque directed along $-\hat{z}$, and exhibits a narrower spectral response around the resonance. Unlike the balanced case $\alpha=\pi/4$, a nonzero torque is observed even at $\theta=\pi/4$, and its magnitude increases as the population is transferred to state $\ket{1}$. Conversely, for $\pi/4<\alpha\leq\pi/2$, the incident vector vortex beam becomes dominated by its left-handed circularly polarized component, with $|\Omega_{L,0}|>|\Omega_{R,0}|$. The corresponding results are shown in Figs.~\ref{Torque spectra lambda alpha variation}(c) and \ref{Torque spectra lambda alpha variation}(f). In this regime, the sign of the torque is reversed: the linear contribution becomes predominantly negative, corresponding to a torque along $-\hat{z}$, whereas the nonlinear contribution becomes predominantly positive, corresponding to a torque along $+\hat{z}$. A finite torque is again observed at $\theta=\pi/4$, with its magnitude increasing as the population accumulates in the state $\ket{2}$. For all values of $\alpha$, the nonlinear contribution remains approximately two orders of magnitude weaker than the linear contribution for $\gamma_d=10^{-3}\gamma$, as evident from the comparison between Figs.~\ref{Torque spectra lambda alpha variation}(a)--(c) and \ref{Torque spectra lambda alpha variation}(d)--(f).

For the tripod configuration, where a strong control field is present, the torque parameter map is shown in Fig.~\ref{Torque spectra tripod alpha variation}. The intensity of the control field is fixed at $|\Omega_C|=\gamma$, while all other parameters are kept identical to those used for the $\Lambda$ configuration in Fig.~\ref{Torque spectra lambda alpha variation}. In the linear regime, Figs.~\ref{Torque spectra tripod alpha variation}(a)–(c) show that the qualitative dependence of the torque on the coherence angle $\theta$ and the input beam polarization parameter $\alpha$ remains similar to that observed in the $\Lambda$ system. However, a key difference arises from the presence of electromagnetically induced transparency: a transparency window forms near resonance ($\Delta/\gamma \approx 0$), which strongly suppresses the torque in this region for all input polarization states, consistent with the behavior discussed in Fig.~\ref{Torque radial pattern in Tripod}. As a result, the torque is redistributed away from the resonance and attains its maximum magnitude for finite detunings around $|\Delta| \approx \gamma$, exhibiting an approximately symmetric response for positive and negative detuning.

\begin{figure*}[!t]
    \centering
    \includegraphics[width=1\textwidth]{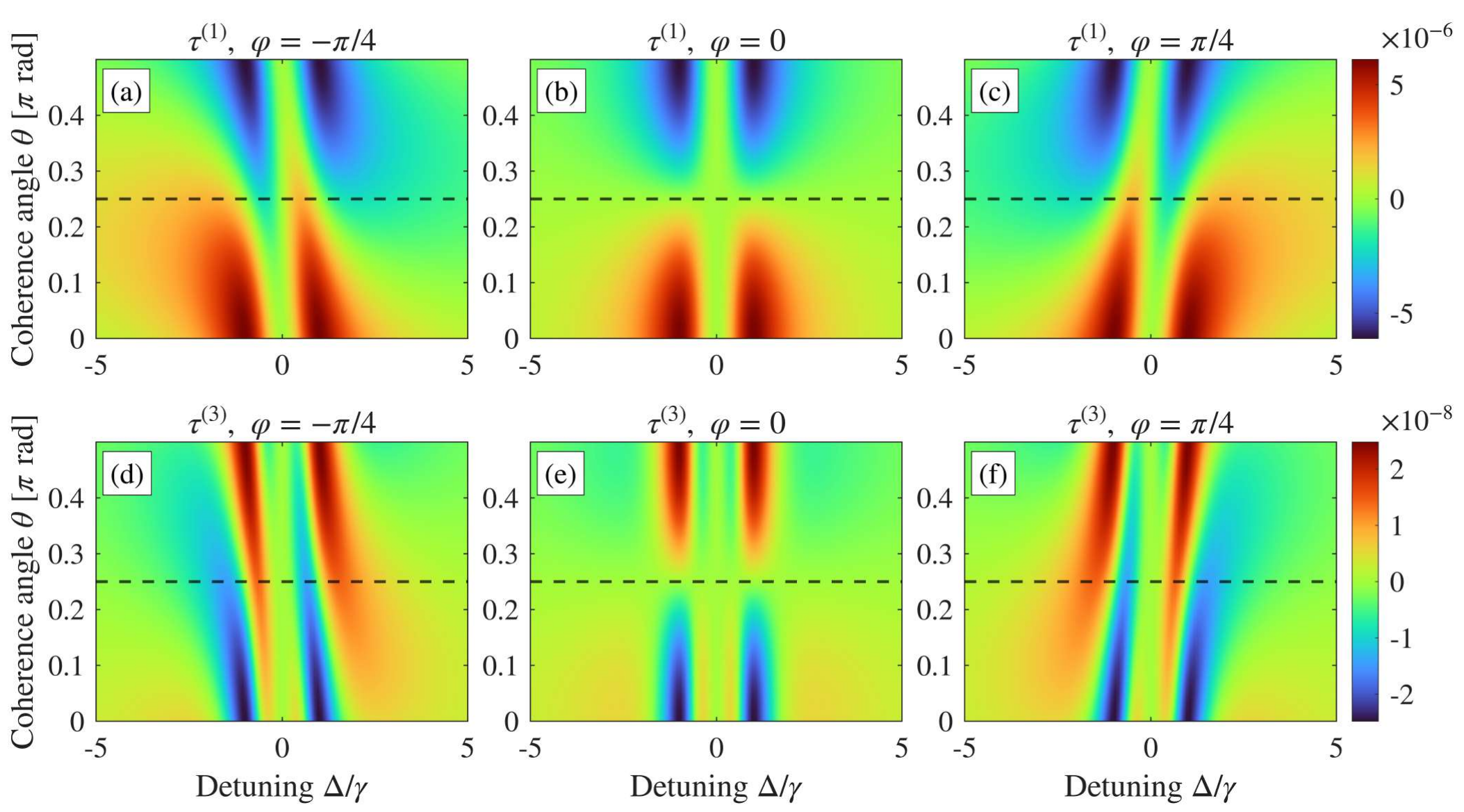}
    \caption{Parameter maps of the optical torque in the tripod configuration for $|l|=1$ at fixed relative amplitude parameter $\alpha=\pi/4$. (a)–(c) Linear contribution to the torque. (d)–(f) Nonlinear contribution to the torque. Three values of the relative phase $\varphi$ are considered: (a),(d) $\varphi=-\pi/4$; (b),(e) $\varphi=0$; (c),(f) $\varphi=\pi/4$. The vertical axis corresponds to the coherence tuning angle $\theta$, where $\theta=\pi/4$ is indicated by a horizontal dashed line. The horizontal axis corresponds to the normalized two-photon detuning. The remaining parameters are $\varepsilon=10^{-2}\gamma$, $\gamma_d=10^{-3}\gamma$, and $|\Omega_C|=\gamma$.}
    \label{Torque spectra tripod varphi}
\end{figure*}

Figures~\ref{Torque spectra tripod alpha variation}(d)–\ref{Torque spectra tripod alpha variation}(f) show the nonlinear torque response in the tripod configuration. Overall, its behavior closely follows that of the nonlinear torque in the $\Lambda$ system. In particular, the nonlinear response exhibits a significantly narrower dependence on detuning $\Delta/\gamma$ compared to the linear contribution, and its sign is reversed relative to the linear torque as a function of the coherence angle $\theta$. For example, for $\alpha=\pi/8$, the linear torque is positive (corresponding to a response along $+\hat{z}$), whereas the nonlinear contribution is negative (along $-\hat{z}$) for $0 \leq \theta < \pi/4$, where the state $\ket{1}$ becomes increasingly populated. In contrast to the $\Lambda$ configuration, the presence of the control field introduces EIT, leading to a suppression of both linear and nonlinear responses near resonance. Consequently, the maxima of the nonlinear torque are shifted away from the resonance and occur around $|\Delta| \approx \gamma$. Finally, the amplitude of the nonlinear torque is reduced by approximately three orders of magnitude compared to the peak value of the linear torque for $\gamma_d = 10^{-3}\gamma$.

Next, we examine the influence of the relative phase $\varphi$ between the right- and left-handed vortex amplitudes on the torque spectra. Throughout this analysis, the amplitude balance is fixed at $\alpha=\pi/4$. Figures~\ref{Torque spectra lambda varphi} and \ref{Torque spectra tripod varphi} present the linear and nonlinear torque responses for the $\Lambda$ and tripod configurations, respectively, for $\varphi=-\pi/4,,0,$ and $\pi/4$.

A nonzero relative phase breaks the symmetry of the torque response in both regimes and in both configurations. In particular, when $\varphi=0$, the torque vanishes at $\theta=\pi/4$ for all detunings, as seen in Figs.~\ref{Torque spectra lambda varphi}(b), \ref{Torque spectra lambda varphi}(e), \ref{Torque spectra tripod varphi}(b), and \ref{Torque spectra tripod varphi}(e), reflecting destructive interference between the two vortex components for equal ground-state populations. In contrast, introducing a phase imbalance leads to finite torque even at $\theta=\pi/4$, provided the system is detuned from resonance. This is illustrated, for example, in the case $\varphi=\pi/4$, where both linear and nonlinear torques become nonzero away from $\Delta/\gamma=0$ in Figs.~\ref{Torque spectra lambda varphi}(c), \ref{Torque spectra lambda varphi}(f), \ref{Torque spectra tripod varphi}(c), and \ref{Torque spectra tripod varphi}(f). The response is antisymmetric with respect to the sign of the relative phase: reversing $\varphi$ reverses the sign of the torque across the detuning axis. This is evident from the comparison between $\varphi=-\pi/4$ and $\varphi=\pi/4$ in both configurations. Overall, the presence of a nonzero relative phase enables both positive and negative torque contributions over different detuning ranges and coherence angles, with a strong sensitivity to interference between the two components of the vector vortex field. This phase-controlled behavior is common to both the $\Lambda$ and tripod configurations, although the tripod case exhibits an additional suppression near resonance due to the anisotropy induced by the control-field.

\section{Concluding remark}
We have theoretically demonstrated an atomic-coherence-sensitive optical torque arising in three-level $\Lambda$ and four-level tripod atom–light configurations, where two coherently prepared ground states (phaseonium) interact with a pair of weak vortex beams carrying opposite orbital angular momenta, $+l$ and $-l$. The spatial phase structure of the optical field is transferred to the atomic medium, resulting in a quantized torque that drives rotational atomic motion along a ring-shaped distribution inherited from the vortex beam profile. Using perturbative steady-state solutions of the optical Bloch equations, we have shown that both linear and nonlinear torque contributions are strongly governed by the initial atomic coherence. In particular, a nonzero torque can arise even when the counter-OAM components have equal amplitudes, due to coherence-induced asymmetry in the atomic response. The torque magnitude is controlled by the beam parameters, including the relative amplitude ratio, relative phase, and detuning, as well as by the initial preparation of the phaseonium . In the tripod configuration, the presence of a strong control field without OAM modifies the torque response and introduces EIT near resonance. This leads to the formation of zero-torque windows around resonance, which persist in both linear and nonlinear regimes. These results demonstrate a mechanism for coherent control of light-induced angular momentum transfer in multi-level atomic media, providing an additional degree of freedom for manipulating rotational atomic dynamics.

Lastly, both the three-level $\Lambda$ configuration and the four-level tripod configuration can be realized experimentally using the hyperfine states of cold $^{87}$Rb atoms. The optical frequencies of the weak vortex pair can be chosen close to the D2 transition line of the hyperfine manifold, for instance, following the $\ket{F=1}\rightarrow\ket{F'=2}$ transition. With this choice of manifolds, the phaseonium lower states, $\ket{1}$ and $\ket{2}$, can be chosen as $\ket{5S_{1/2},F=1,m_f=-1}$ and $\ket{5S_{1/2},F=1,m_f=+1}$, respectively, whereas the excited state can be chosen as $\ket{5P_{3/2},F'=1,m_f'=0}$. The phaseonium lower states can be prepared using a coherent population transfer by laser radiation, such as stimulated Raman adiabatic passage (STIRAP), by employing additional pump and Stokes lasers with controlled pulse delay and intensity variation. This allows the population to be adiabatically transferred between the ground states, $\ket{2}\rightarrow\ket{1}$, while avoiding losses caused by significant population transfer to the excited state, $\ket{0}$ \cite{bergmann2017}. In the tripod configuration, a state from a different ground-state manifold of the D2 transition line can be exploited as the third unoccupied ground state, $\ket{3}$; for example, $\ket{5S_{1/2},F=2,m_f=\pm1}$ can be used, where an OAM-free control beam with $\hat{\sigma}_{\pm}$ polarization couples the $\ket{3}\rightarrow\ket{0}$ transition.

\begin{acknowledgments}
This project has received funding from the Research Council of Lithuania (LMTLT), agreement No. S-ITP-24-6. D.P.P. gratefully acknowledges the support of the Erasmus+: Erasmus Mundus programme of the European Union under Convention $\mathrm{n^o}$ 101128124 — EUROPHOTONICS — ERASMUS-EDU-2023-PEX-EMJMMOB. 
\end{acknowledgments}

\bibliography{apssamp.bib}

@PREAMBLE{
 "\providecommand{\noopsort}[1]{}" 
 # "\providecommand{\singleletter}[1]{#1}%" 
}

@article{Babiker2010,
  title = {Light-induced torque for the generation of persistent current flow in atomic gas Bose-Einstein condensates},
  author = {Lembessis, V. E. and Babiker, M.},
  journal = {Phys. Rev. A},
  volume = {82},
  issue = {5},
  pages = {051402(R)},
  numpages = {4},
  year = {2010},
  month = {Nov},
  publisher = {American Physical Society},
  doi = {10.1103/PhysRevA.82.051402},
  url = {https://link.aps.org/doi/10.1103/PhysRevA.82.051402}
}

@article{Babiker1994,
  title = {Light-induced Torque on Moving Atoms},
  author = {Babiker, M. and Power, W. L. and Allen, L.},
  journal = {Phys. Rev. Lett.},
  volume = {73},
  issue = {9},
  pages = {1239--1242},
  numpages = {0},
  year = {1994},
  month = {Aug},
  publisher = {American Physical Society},
  doi = {10.1103/PhysRevLett.73.1239},
  url = {https://link.aps.org/doi/10.1103/PhysRevLett.73.1239}
}

@article{Phillips2006,
  title = {Quantized Rotation of Atoms from Photons with Orbital Angular Momentum},
  author = {Andersen, M. F. and Ryu, C. and Clad\'e, Pierre and Natarajan, Vasant and Vaziri, A. and Helmerson, K. and Phillips, W. D.},
  journal = {Phys. Rev. Lett.},
  volume = {97},
  issue = {17},
  pages = {170406},
  numpages = {4},
  year = {2006},
  month = {Oct},
  publisher = {American Physical Society},
  doi = {10.1103/PhysRevLett.97.170406},
  url = {https://link.aps.org/doi/10.1103/PhysRevLett.97.170406}
}

@article{Cook1979,
  title = {Atomic motion in resonant radiation: An application of Ehrenfest's theorem},
  author = {Cook, R. J.},
  journal = {Phys. Rev. A},
  volume = {20},
  issue = {1},
  pages = {224--228},
  numpages = {0},
  year = {1979},
  month = {Jul},
  publisher = {American Physical Society},
  doi = {10.1103/PhysRevA.20.224},
  url = {https://link.aps.org/doi/10.1103/PhysRevA.20.224}
}

@article{Permana2026,
title = {Propagation of optical vector vortices of slow light in a coherently prepared tripod configuration},
journal = {Chaos, Solitons \& Fractals},
volume = {209},
pages = {118518},
year = {2026},
issn = {0960-0779},
doi = {https://doi.org/10.1016/j.chaos.2026.118518},
url = {https://www.sciencedirect.com/science/article/pii/S0960077926006594},
author = {Dharma P. Permana and Mazena Mackoit Sinkevičienė and Julius Ruseckas and Hamid R. Hamedi}
}

@article{Hamid2025,
  title = {Coherent phase control of orbital-angular-momentum light-induced torque in a double-tripod atom-light coupling scheme},
  author = {Hamedi, Hamid R. and Kudriašov, Viačeslav and Sinkevičienė, Mažena Mackoit and Ruseckas, Julius},
  journal = {Phys. Rev. A},
  volume = {112},
  issue = {6},
  pages = {063720},
  numpages = {10},
  year = {2025},
  month = {Dec},
  publisher = {American Physical Society},
  doi = {10.1103/nbx8-l3v4},
  url = {https://link.aps.org/doi/10.1103/nbx8-l3v4}
}

@article{Peters2022,
author = {Thorsten Peters and Yi-Hsin Chen and Jian-Siung Wang and Yen-Wei Lin and Ite A. Yu},
journal = {Opt. Lett.},
number = {2},
pages = {151--153},
publisher = {Optica Publishing Group},
title = {Observation of phase variation within stationary light pulses inside a cold atomic medium},
volume = {35},
month = {Jan},
year = {2010},
url = {https://opg.optica.org/ol/abstract.cfm?URI=ol-35-2-151},
doi = {10.1364/OL.35.000151},
}

@article{Kim2022,
  title = {Simultaneous Trapping of Two Optical Pulses in an Atomic Ensemble as Stationary Light Pulses},
  author = {Kim, U-Shin and Kim, Yoon-Ho},
  journal = {Phys. Rev. Lett.},
  volume = {129},
  issue = {9},
  pages = {093601},
  numpages = {6},
  year = {2022},
  month = {Aug},
  publisher = {American Physical Society},
  doi = {10.1103/PhysRevLett.129.093601},
  url = {https://link.aps.org/doi/10.1103/PhysRevLett.129.093601}
}

@article{Otterbach2010,
  title = {Effective Magnetic Fields for Stationary Light},
  author = {Otterbach, J. and Ruseckas, J. and Unanyan, R. G. and Juzeli\ifmmode \bar{u}\else \={u}\fi{}nas, G. and Fleischhauer, M.},
  journal = {Phys. Rev. Lett.},
  volume = {104},
  issue = {3},
  pages = {033903},
  numpages = {4},
  year = {2010},
  month = {Jan},
  publisher = {American Physical Society},
  doi = {10.1103/PhysRevLett.104.033903},
  url = {https://link.aps.org/doi/10.1103/PhysRevLett.104.033903}
}

@Article{Hau1999,
author={Hau, Lene Vestergaard
and Harris, S. E.
and Dutton, Zachary
and Behroozi, Cyrus H.},
title={Light speed reduction to 17 metres per second in an ultracold atomic gas},
journal={Nature},
year={1999},
month={Feb},
day={01},
volume={397},
number={6720},
pages={594-598},
issn={1476-4687},
doi={10.1038/17561},
url={https://doi.org/10.1038/17561}
}

@article{Rebic2004,
  title = {Polarization phase gate with a tripod atomic system},
  author = {Rebi\ifmmode \acute{c}\else \'{c}\fi{}, S. and Vitali, D. and Ottaviani, C. and Tombesi, P. and Artoni, M. and Cataliotti, F. and Corbal\'an, R.},
  journal = {Phys. Rev. A},
  volume = {70},
  issue = {3},
  pages = {032317},
  numpages = {8},
  year = {2004},
  month = {Sep},
  publisher = {American Physical Society},
  doi = {10.1103/PhysRevA.70.032317},
  url = {https://link.aps.org/doi/10.1103/PhysRevA.70.032317}
}

@article{Permana2025,
  title = {Spin-orbit coupling of optical vector vortices in coherently prepared media},
  author = {Permana, Dharma P. and Sinkevi\ifmmode \check{c}\else \v{c}\fi{}ien\ifmmode \dot{e}\else \.{e}\fi{}, Mazena Mackoit and Ruseckas, Julius and Hamedi, Hamid R.},
  journal = {Phys. Rev. A},
  volume = {113},
  issue = {4},
  pages = {043705},
  numpages = {12},
  year = {2026},
  month = {Apr},
  publisher = {American Physical Society},
  doi = {10.1103/trtn-m1tw},
  url = {https://link.aps.org/doi/10.1103/trtn-m1tw}
}

@book{scully.book1997, place={Cambridge}, title={Quantum Optics}, publisher={Cambridge University Press}, author={Scully, Marlan O. and Zubairy, M. Suhail}, year={1997}}

@article{Hau2001,
author={Liu, Chien
and Dutton, Zachary
and Behroozi, Cyrus H.
and Hau, Lene Vestergaard},
title={Observation of coherent optical information storage in an atomic medium using halted light pulses},
journal={Nature},
year={2001},
month={Jan},
day={01},
volume={409},
number={6819},
pages={490-493},
issn={1476-4687},
doi={10.1038/35054017},
url={https://doi.org/10.1038/35054017}
}

@article{lukin2001,
  title = {Storage of Light in Atomic Vapor},
  author = {Phillips, D. F. and Fleischhauer, A. and Mair, A. and Walsworth, R. L. and Lukin, M. D.},
  journal = {Phys. Rev. Lett.},
  volume = {86},
  issue = {5},
  pages = {783--786},
  numpages = {0},
  year = {2001},
  month = {Jan},
  publisher = {American Physical Society},
  doi = {10.1103/PhysRevLett.86.783},
  url = {https://link.aps.org/doi/10.1103/PhysRevLett.86.783}
}

@article{bergmann2017,
  title = {Stimulated {R}aman adiabatic passage in physics, chemistry, and beyond},
  author = {Vitanov, Nikolay V. and Rangelov, Andon A. and Shore, Bruce W. and Bergmann, Klaas},
  journal = {Rev. Mod. Phys.},
  volume = {89},
  issue = {1},
  pages = {015006},
  numpages = {66},
  year = {2017},
  month = {Mar},
  publisher = {American Physical Society},
  doi = {10.1103/RevModPhys.89.015006},
  url = {https://link.aps.org/doi/10.1103/RevModPhys.89.015006}
}

@article{rosales2018, title={A review of complex vector light fields and their applications}, volume={20}, DOI={10.1088/2040-8986/aaeb7d}, number={12}, journal={Journal of Optics}, author={Rosales-Guzmán, Carmelo and Ndagano, Bienvenu and Forbes, Andrew}, year={2018}, month={Nov}, pages={123001}}

@article{guo2008,
  title = {Entanglement of the orbital angular momentum states of the photon pairs generated in a hot atomic ensemble},
  author = {Chen, Qun-Feng and Shi, Bao-Sen and Zhang, Yong-Sheng and Guo, Guang-Can},
  journal = {Phys. Rev. A},
  volume = {78},
  issue = {5},
  pages = {053810},
  numpages = {4},
  year = {2008},
  month = {Nov},
  publisher = {American Physical Society},
  doi = {10.1103/PhysRevA.78.053810},
  url = {https://link.aps.org/doi/10.1103/PhysRevA.78.053810}
}

@article{ruseckas2013,
  title = {Transfer of orbital angular momentum of light using two-component slow light},
  author = {Ruseckas, Julius and Kudriašov, Viačeslav and Yu, Ite A. and Juzeliūnas, Gediminas},
  journal = {Phys. Rev. A},
  volume = {87},
  issue = {5},
  pages = {053840},
  numpages = {6},
  year = {2013},
  month = {May},
  publisher = {American Physical Society},
  doi = {10.1103/PhysRevA.87.053840},
  url = {https://link.aps.org/doi/10.1103/PhysRevA.87.053840}
}

@article{tarak2017,
  title = {Phase-induced transparency-mediated structured-beam generation in a closed-loop tripod configuration},
  author = {Sharma, Sandeep and Dey, Tarak N.},
  journal = {Phys. Rev. A},
  volume = {96},
  issue = {3},
  pages = {033811},
  numpages = {8},
  year = {2017},
  month = {Sep},
  publisher = {American Physical Society},
  doi = {10.1103/PhysRevA.96.033811},
  url = {https://link.aps.org/doi/10.1103/PhysRevA.96.033811}
}

@article{radwell2017,
  title = {Spatially Dependent Electromagnetically Induced Transparency},
  author = {Radwell, N. and Clark, T. W. and Piccirillo, B. and Barnett, S. M. and Franke-Arnold, S.},
  journal = {Phys. Rev. Lett.},
  volume = {114},
  issue = {12},
  pages = {123603},
  numpages = {5},
  year = {2015},
  month = {Mar},
  publisher = {American Physical Society},
  doi = {10.1103/PhysRevLett.114.123603},
  url = {https://link.aps.org/doi/10.1103/PhysRevLett.114.123603}
}

@article{hamedi2015,
  title = {Phase-sensitive {K}err nonlinearity for closed-loop quantum systems},
  author = {Hamedi, H. R. and Juzeli\ifmmode \bar{u}\else \={u}\fi{}nas, G.},
  journal = {Phys. Rev. A},
  volume = {91},
  issue = {5},
  pages = {053823},
  numpages = {13},
  year = {2015},
  month = {May},
  publisher = {American Physical Society},
  doi = {10.1103/PhysRevA.91.053823},
  url = {https://link.aps.org/doi/10.1103/PhysRevA.91.053823}
}

@article{gong2006,
  title = {Enhancing {K}err nonlinearity via spontaneously generated coherence},
  author = {Niu, Yueping and Gong, Shangqing},
  journal = {Phys. Rev. A},
  volume = {73},
  issue = {5},
  pages = {053811},
  numpages = {6},
  year = {2006},
  month = {May},
  publisher = {American Physical Society},
  doi = {10.1103/PhysRevA.73.053811},
  url = {https://link.aps.org/doi/10.1103/PhysRevA.73.053811}
}

@article{minxiao2001,
  title = {Enhanced {K}err Nonlinearity via Atomic Coherence in a Three-Level Atomic System},
  author = {Wang, Hai and Goorskey, David and Xiao, Min},
  journal = {Phys. Rev. Lett.},
  volume = {87},
  issue = {7},
  pages = {073601},
  numpages = {4},
  year = {2001},
  month = {Jul},
  publisher = {American Physical Society},
  doi = {10.1103/PhysRevLett.87.073601},
  url = {https://link.aps.org/doi/10.1103/PhysRevLett.87.073601}
}

@article{juzeliunas2004,
  title = {Slow Light in Degenerate Fermi Gases},
  author = {Juzeli\ifmmode \bar{u}\else \={u}\fi{}nas, G. and \"Ohberg, P.},
  journal = {Phys. Rev. Lett.},
  volume = {93},
  issue = {3},
  pages = {033602},
  numpages = {4},
  year = {2004},
  month = {Jul},
  publisher = {American Physical Society},
  doi = {10.1103/PhysRevLett.93.033602},
  url = {https://link.aps.org/doi/10.1103/PhysRevLett.93.033602}
}

@article{paspalakis2002,
  title = {Electromagnetically induced transparency and controlled group velocity in a multilevel system},
  author = {Paspalakis, E. and Knight, P. L.},
  journal = {Phys. Rev. A},
  volume = {66},
  issue = {1},
  pages = {015802},
  numpages = {4},
  year = {2002},
  month = {Jul},
  publisher = {American Physical Society},
  doi = {10.1103/PhysRevA.66.015802},
  url = {https://link.aps.org/doi/10.1103/PhysRevA.66.015802}
}

@article{fleischhauer2000,
  title = {Dark-State Polaritons in Electromagnetically Induced Transparency},
  author = {Fleischhauer, M. and Lukin, M. D.},
  journal = {Phys. Rev. Lett.},
  volume = {84},
  issue = {22},
  pages = {5094--5097},
  numpages = {0},
  year = {2000},
  month = {May},
  publisher = {American Physical Society},
  doi = {10.1103/PhysRevLett.84.5094},
  url = {https://link.aps.org/doi/10.1103/PhysRevLett.84.5094}
}

@article{boller1991,
  title = {Observation of electromagnetically induced transparency},
  author = {Boller, K.-J. and Imamo\ifmmode \breve{g}\else \u{g}\fi{}lu, A. and Harris, S. E.},
  journal = {Phys. Rev. Lett.},
  volume = {66},
  issue = {20},
  pages = {2593--2596},
  numpages = {0},
  year = {1991},
  month = {May},
  publisher = {American Physical Society},
  doi = {10.1103/PhysRevLett.66.2593},
  url = {https://link.aps.org/doi/10.1103/PhysRevLett.66.2593}
}

@article{zhang2010,
  title = {Implementation of one-dimensional quantum walks on spin-orbital angular momentum space of photons},
  author = {Zhang, Pei and Liu, Bi-Heng and Liu, Rui-Feng and Li, Hong-Rong and Li, Fu-Li and Guo, Guang-Can},
  journal = {Phys. Rev. A},
  volume = {81},
  issue = {5},
  pages = {052322},
  numpages = {4},
  year = {2010},
  month = {May},
  publisher = {American Physical Society},
  doi = {10.1103/PhysRevA.81.052322},
  url = {https://link.aps.org/doi/10.1103/PhysRevA.81.052322}
}

@article{allen1992,
  title = {Orbital angular momentum of light and the transformation of {L}aguerre-{G}aussian laser modes},
  author = {Allen, L. and Beijersbergen, M. W. and Spreeuw, R. J. C. and Woerdman, J. P.},
  journal = {Phys. Rev. A},
  volume = {45},
  issue = {11},
  pages = {8185--8189},
  numpages = {0},
  year = {1992},
  month = {Jun},
  publisher = {American Physical Society},
  doi = {10.1103/PhysRevA.45.8185},
  url = {https://link.aps.org/doi/10.1103/PhysRevA.45.8185}
}

@article{coullet1989,
title = {Optical vortices},
journal = {Optics Communications},
volume = {73},
number = {5},
pages = {403-408},
year = {1989},
issn = {0030-4018},
doi = {https://doi.org/10.1016/0030-4018(89)90180-6},
url = {https://www.sciencedirect.com/science/article/pii/0030401889901806},
author = {P. Coullet and L. Gil and F. Rocca}
}

@article{Marangos.RMP2005,
  title = {Electromagnetically induced transparency: Optics in coherent media},
  author = {Fleischhauer, Michael and Imamoglu, Atac and Marangos, Jonathan P.},
  journal = {Rev. Mod. Phys.},
  volume = {77},
  issue = {2},
  pages = {633--673},
  numpages = {0},
  year = {2005},
  month = {Jul},
  publisher = {American Physical Society},
  doi = {10.1103/RevModPhys.77.633},
  url = {https://link.aps.org/doi/10.1103/RevModPhys.77.633}
}

@article{Kudriasov.OE2025,
author = {Viačeslav Kudriašov and Mazena Mackoit Sinkevičienė and Nilamoni Daloi and Julius Ruseckas and Tarak N. Dey and Sonja Franke-Arnold and Hamid R. Hamedi},
journal = {Opt. Express},
number = {19},
pages = {40931--40947},
publisher = {Optica Publishing Group},
title = {Propagation of optical vector and scalar vortices in an atomic medium with closed-loop tripod configuration},
volume = {33},
month = {Sep},
year = {2025},
url = {https://opg.optica.org/oe/abstract.cfm?URI=oe-33-19-40931},
doi = {10.1364/OE.566363},
}

@article{Hamid.PRA2019,
  title = {Transfer of optical vortices in coherently prepared media},
  author = {Hamedi, Hamid Reza and Ruseckas, Julius and Paspalakis, Emmanuel and Juzeli\ifmmode \bar{u}\else \={u}\fi{}nas, Gediminas},
  journal = {Phys. Rev. A},
  volume = {99},
  issue = {3},
  pages = {033812},
  numpages = {9},
  year = {2019},
  month = {Mar},
  publisher = {American Physical Society},
  doi = {10.1103/PhysRevA.99.033812},
  url = {https://link.aps.org/doi/10.1103/PhysRevA.99.033812}
}

\end{document}